# Building spin-1/2 antiferromagnetic Heisenberg chains with diaza-nanographenes

Xiaoshuai Fu[1,2,†], Li Huang[1,2,7,†]*, Kun Liu[3,†], João C. G. Henriques[5,9], Yixuan Gao[1,2], Xianghe Han[1,2], Hui Chen[1,2], Yan Wang[1,2], Carlos-Andres Palma[1,2], Zhihai Cheng[6], Xiao Lin[2,1], Shixuan Du[1,2], Ji Ma[4,8]*, Joaquín Fernández-Rossier[5]*, Xinliang Feng[3,8], Hong-Jun Gao[1,2,7]

[1] Beijing National Center for Condensed Matter Physics and Institute of Physics, Chinese Academy of Sciences, Beijing 100190, China.

[2] School of Physical Sciences, University of Chinese Academy of Sciences, Beijing 100190, China

[3] Center for Advancing Electronics Dresden (cfaed) & Faculty of Chemistry and Food Chemistry, Technische Universität Dresden, D-01069 Dresden, Germany

[4] College of Materials Science and Optoelectronic Technology, University of Chinese Academy of Sciences, Beijing 100190, China

[5] International Iberian Nanotechnology Laboratory, 4715-310 Braga, Portugal

[6] Beijing Key Laboratory of Optoelectronic Functional Materials & Micro-Nano Devices, Department of Physics, Renmin University of China, Beijing 100872, China

[7] Hefei National Laboratory, Hefei, Anhui 230088, China

[8] Max Planck Institute of Microstructure Physics, Weinberg 2, 06120 Halle, Germany

[9] Universidade de Santiago de Compostela, 15782 Santiago de Compostela, Spain

* Corresponding author. Email: lhuang@iphy.ac.cn; maji@ucas.ac.cn; joaquin.fernandez-rossier@inl.int

† These authors contributed equally to this work

**Abstract**

Understanding and engineering the coupling of spins in nanomaterials is of central importance for designing novel devices. Graphene nanostructures with π-magnetism offer a chemically tunable platform to explore quantum magnetic interactions. However, realizing spin chains bearing controlled odd-even effects with suitable nanographene systems is challenging. Here, we demonstrate the successful on-surface synthesis of spin-1/2 antiferromagnetic Heisenberg chains with parity-dependent magnetization based on antiaromatic diaza-hexa-*peri*-hexabenzocoronene (diaza-HBC) units. Using distinct synthetic strategies, two types of spin chains with different terminals were synthesized, both exhibiting a robust odd-even effect on the spin coupling along the chain. Combined investigations using scanning tunneling microscopy, non-contact atomic force microscopy, density functional theory calculations, and quantum spin models confirmed the structures of the diaza-HBC chains and revealed their magnetic properties, which has an $S = 1/2$ spin per unit through electron donation from the diaza-HBC core to the Au(111) substrate. Gapped excitations were observed in even-numbered chains, while enhanced Kondo resonance emerged in odd-numbered units of odd-numbered chains due to the redistribution of the unpaired spin along the chain. Our findings provide an effective strategy to construct nanographene spin chains and unveil the odd-even effect in their magnetic properties, offering potential applications in nanoscale spintronics.

**Introduction**

The magnetic order of spin-polarized π-electrons in graphene nanostructures has received significant attention due to its potential applications in spintronics and quantum informatics.[1-7] The delocalization of π-electrons allows the engineering of quantum nanomagnet architectures with coupled spins based on atomically defined structures and sizes.[8-12] Thus far, most nanographenes hosting π-magnetism have a non-Kekulé structure, which is highly reactive and unstable, making them challenging to synthesize through chemical reactions in solution. In recent years, bottom-up on-surface synthesis, starting from predesigned precursors, has been developed into a powerful tool in the constructions of atomically precise spin-polarized graphene nanostructures, such as triangulenes, rhombenes, Clar's goblet, and zigzag-edged graphene nanoribbons.[13-23] These nanostructures have a surplus of carbon atoms on sublattice A over sublattice B, leading to a spin-polarized state at the Fermi level.[24] Construction of antiferromagnetic Heisenberg spin chains is of vital importance to investigate magnetism with strong electron correlations.[25-27] In finite one-dimensional (1D) Heisenberg chains, their length and symmetry have a strong influence on their magnetic behavior,[28] such as the existence of an odd-even effect on the ground state of atomic chains on ferromagnetic substrates.[29-31] Magnetic chains with $S = 1$ triangulene or porphyrin units, which exhibit an antiferromagnetic order and fractional edge excitations, have been achieved on the metal surface and proved to have a topological Haldane spin chain phase.[32,33] However, spin $S = 1/2$ antiferromagnetic Heisenberg chains have yet to be thoroughly investigated due to a lack of suitable building units.

Heteroatom doping is another effective strategy for inducing magnetism in graphene nanostructures.[34] Substitution of trigonal planar C atoms with graphitic N (B) atoms would introduce a single surplus (deficit) π-electron per dopant, thus forming an unpaired spin. Graphene nanoribbons with periodic $N_2$ or $B_2$ dopants have been synthesized on the Au(111) surface and revealed the signature of a Kondo resonance emerging from the interaction between the $S = 1/2$ spin center at each dopant site with the itinerant electrons in the Au substrate.[35-38] Reactions based on polycyclic aromatic azomethine ylide (PAMY) provide a unique entry with high versatility to N-doped nanographenes.[39-42] The neutral form of PAMY, possessing a zigzag edge embedded with an N dopant and two *ortho sp*$^3$ C atoms, also exhibits high selective reactivity through C-H activation to synthesize designed graphene nanostructures.[43] Therefore, PAMY chemistry can serve as a promising strategy to construct collective π-magnetism in N-doped graphene nanostructures.

Here, we report the successful synthesis of two types of 1D spin-1/2 spin chains (**chain 1** and **chain 2**) on Au(111) surface, in which their building unit is based on antiaromatic diaza-hexa-*peri*-hexabenzocoronene (diaza-HBC). **Chain 1** is purely comprised of diaza-HBC units yielded through C-H activation and subsequent Ullmann polymerization; while **chain 2**, with lengths of up to 8 units, has two terminals attached to the ends of the diaza-HBC chain by carbon-carbon single bond, synthesized through C-H activation without any halogen elements involved in the polymerization process. Despite their difference in structures, **chain 1** and **chain 2** with the same unit numbers exhibit comparable intermolecular antiferromagnetic exchange as provided by parity-dependent inelastic electron tunnel spectroscopy. The magnetism is generated by electron donation from each diaza dopant center to the substrate, as confirmed by density functional theory (DFT) calculations. For even-numbered chains, inelastic electron tunneling spectroscopy manifests finite energy inelastic excitations, while the odd-numbered chains feature a zero-bias Kondo resonance, whose intensity is modulated across different units. This odd-even effect is robust on both **chain 1** and **chain 2**, regardless of the terminal structures. These observations are naturally understood in terms of the $S = 1/2$ Heisenberg model with antiferromagnetic coupling, where each diaza-HBC is one site in the spin chain, for which even-numbered (odd-numbered) chains have a ground state singlet (doublet) with $S = 0$ ($S = 1/2$). For odd-numbered chains, the model predicts a larger local average magnetization $<S_z>$ in the odd-numbered sites, leading to an enhancement of the Kondo resonance in these sites. Our work demonstrates the first nanographene spin chains that fit the $S = 1/2$ antiferromagnetic Heisenberg chain model, offering the potential for the development of novel spin-based technologies.

**Results and discussion**

Precursor **1**, bearing a graphitic N heteroatom embedded in a zigzag edge, undergoes surface-catalyzed selective C-H activation at the two *ortho* $sp^3$ C atoms at 370°C on Au(111), as depicted in Fig. 1a. These *ortho* $sp^3$ C atoms dissociate to yield $sp^2$ C atoms, leading to diradical type **4** or azomethine ylide type **4'**. These reactive intermediates homocouple through their N-embedded zigzag edges, and form diaza-HBC monomers (Fig. 1b) by subsequent surface-assisted cyclodedydrogenation along with the formation of two C-C single bonds and two C=C double bonds, which are labeled in red in the diaza-HBC chemical structure in Figure 1a. Scanning diaza-HBC monomers at +3 V (Fig. 1c) clearly presents the six-fold flower-like feature, which is identical to the monomers synthesized based on the salt form of PAMYs.[39] The nc-AFM image in Fig. 1e resolved the chemical structure of the product using a CO-functionalized tip, showing the N-doped antiaromatic center a little darker than the carbon skeleton, consistent with previous nc-AFM images on N dopants.[39, 44] Compared with the reaction starting from the salt form of PAMY,[39] our reaction has an excellent diaza-HBC yield of ~84% (Fig. 1d and Fig. S1), which is essential for designing precursors that enable further polymerization to form antiaromatic chains.

The dI/dV spectrum on diaza-HBC (blue) in Fig. 1f exhibits a prominent peak at the Fermi level, which was taken at the center of the diaza-HBC, as denoted by the blue dot in the inset figure. In the dI/dV map in Fig. 1g, this zero-bias peak spatially distributes at the center and along the periphery edges parallel the $N_2$ dopant, showing a delocalized feature over the molecule. DFT calculated electron density difference of diaza-HBC adsorbed on Au(111) shows charge density depletion at the antiaromatic diaza doping center and the periphery of the diaza-HBC, as indicated by the cyan contour regions in Fig. 1h; while there is charge density accumulation on the Au(111) substrate underneath the doping center, as denoted by the yellow contour in the side view image in the upper panel of Fig. 1i. The electron density difference curve (lower panel of Fig. 1i) demonstrates that diaza-HBC has a radical cation formed on Au(111) with almost one electron transferred to the substrate, giving a $S = 1/2$ ground state and spin-polarized orbitals near the Fermi level (Fig. 2j). This is the first example to make $S = 1/2$ from the antiaromatic polycyclic hydrocarbons. Therefore, the zero-bias peak in dI/dV spectrum is a Kondo resonance when the net spin on the diaza-HBC$^+$ unit interacts with the conduction electrons in Au(111) substrate.

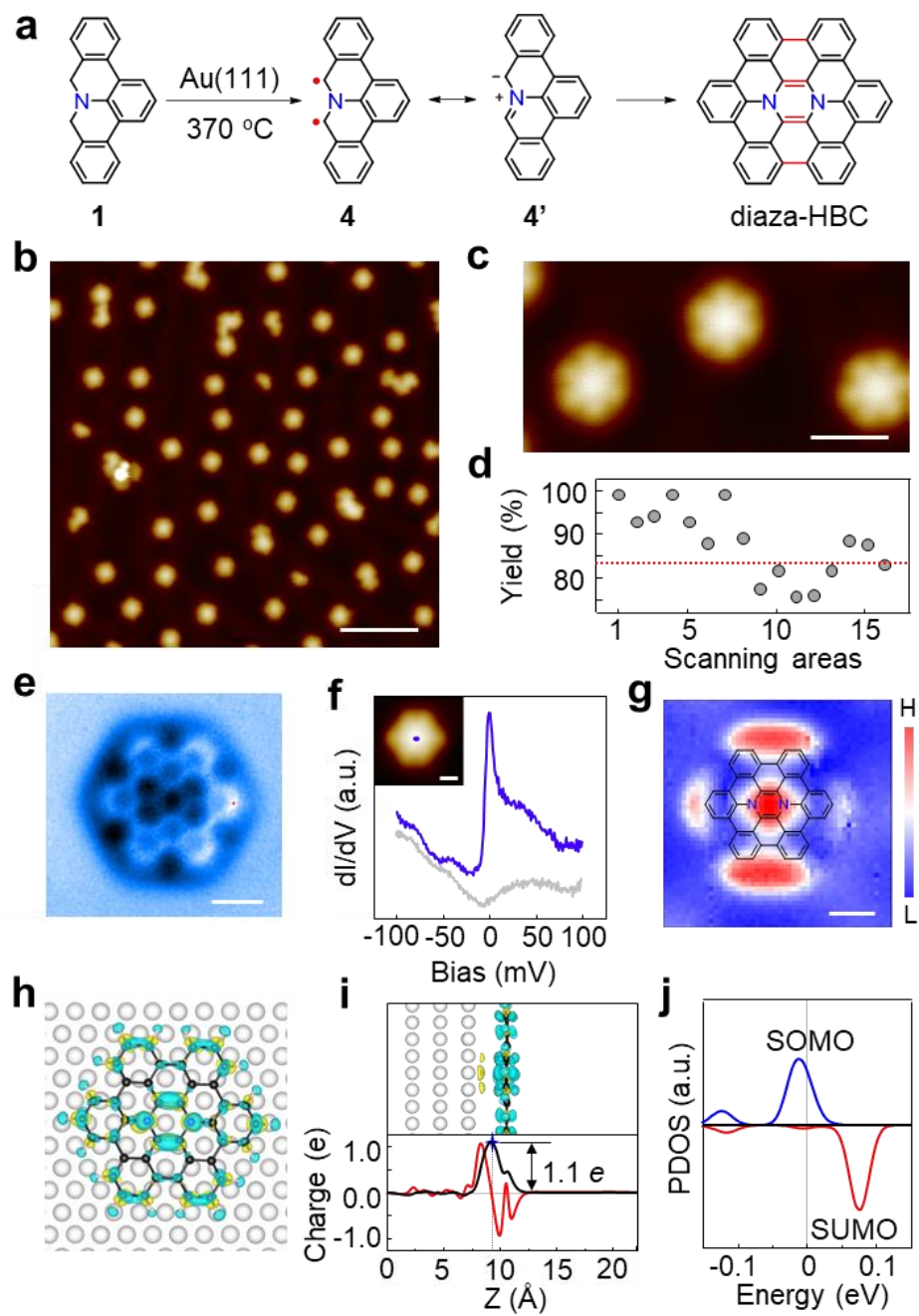

**Fig. 1 On-surface synthesis and Kondo resonance on diaza-HBC monomers.** (**a**) Scheme of the synthesis of diaza-HBC monomers starting from precursor **1**. (**b**) Large-are STM image of diaza-HBC after annealing precursor **1** on Au(111) at 370 ℃ (V = -0.1 V, I = 10 pA). (**c**) Zoom-in STM image of diaza-HBC monomers on Au(111) (V = 3 V, I = 10 pA). (**d**) The statistics on the yield rate of diaza-HBC monomers over 16 different scanning areas. The corresponding STM images used for the statistics are presented in Fig. S1. (**e**) Chemical-bond-resolved nc-AFM image of diaza-HBC (oscillation amplitude $A_{OSC}$ = 100 pm). The image rotates to the same orientation as the chemical structure of diaza-HBC in (a). (**f**) dI/dV spectra taken at the center of a diaza-HBC monomer, as marked by the blue dot in the inset (blue), and on bare Au(111) (grey). Modulation amplitudes $V_{AC}$ = 1 mV. The inset is the STM image of a diaza-HBC monomer on Au(111). (V = -0.5 V, I = 100 pA) (**g**) dI/dV map taken at 0 V of the same diaza-HBC monomer in the inset of (**f**), with the chemical structure superimposed. (V = -50 mV, $V_{AC}$ = 5 mV) (**h**) DFT calculated electron density difference on diaza-

HBC/Au(111). The cyan contour shows regions of charge density depletion, while the yellow contour shows regions of charge density accumulation. (**i**) Upper panel: the side view of the diaza-HBC and the Au(111) substrate showing charge transfer from diaza-HBC to Au(111) with the same color code as in (h). The charge density difference is defined as $\Delta \rho = \rho(\text{tot}) - \rho(\text{substrate}) - \rho(\text{molecule})$, where $\rho(\text{tot})$ is the total charge density, $\rho(\text{substrate})$ is the Au(111) substrate charge density, and $\rho(\text{molecule})$ is the charge density of diaza-HBC monomer. Lower panel: the DFT calculated in-plane electron density difference curve. The red curve represents the in-plan charge density difference, and the black curve is the integration of the in-plan charge density difference along z direction. (**j**) Projected density of states (PDOS) for the diaza-HBC on Au(111) substrate. The blue and red curves denote PDOS with different spins. Scale bars: **b** 6 nm; **c** 2 nm; **e**, **f**, and **g** 0.5 nm.

To integrate diaza-HBC monomers into chains, two kinds of precursors (**2** and **3**) were designed and synthesized (details in Supplementary Information Scheme 1 and Fig. S1-S6). As depicted in Fig. 2a, dehydrogenation at the two *ortho sp*$^3$ C atoms of precursor **2** transforms it into diradical type **5** or its resonant azomethine ylide form on Au(111) at an annealing temperature of 370°C. The reactive diradical further undergoes homocoupling into **6**. Then dechlorination, C-C coupling, and cyclodehydrogenation occur to yield **chain 1** comprised of pure diaza-HBC units. A small amount of side products was also observed on the surface, which mainly results from the coupling between the zigzag edge and the dechlorinated carbon atom, indicating that dechlorination may occur at a comparable temperature of *ortho sp*$^3$ C-H activation. The large-area STM image of **chain 1** in Fig. 2b exhibits that the majority of the products are diaza-HBC monomers, as denoted by the blue dashed circle, and **chain 1** with various lengths can be found on the surface, as indicated by the blue dashed boxes. The nc-AFM image of an n = 4 **chain 1** is displayed in Fig. 2d. The STM and nc-AFM images of **chain 1** with n = 2-6 are shown in Fig. S8.

Precursor **3** has a pair of mirror-reversed units of **1**, connected through a single bond, exhibiting the reactive *ortho sp*$^3$ C atoms on the two outer zigzag edges. Precursor **3** undergoes homocoupling through tetraradical intermediate **7** when annealed on Au(111) at 370 °C, resulting in the formation of **chain 2** after C-H activation, C-C homocoupling, and cyclodehydrogenation. Compared to **chain 1**, **chain 2** has two terminal units attached to both ends of the diaza-HBC chain, as shown in Fig. 2a. A large-area STM image of **chain 2** in Fig. 2c shows chains grown along different directions, with some of the chains labeled by blue dashed boxes for clarity. We have synthesized **chain 2** with lengths of up to 8 units (Fig. S9).

Figure 2e shows the nc-AFM image of an n = 6 **chain 2**, which has different contrast along the zigzag edges of both terminal units. The brighter contrast, as denoted by the red arrows in Fig. 2e, indicates the existence of out-of-plane C-H bonds, which implies that these C atoms are $sp^3$ hybridized. The darker contrast (the yellow arrows in Fig. 2e) indicates that the $sp^3$ C atom has dehydrogenated into planar $sp^2$ C with a radical. This methylene radical type of terminal unit, as depicted in the chemical structure in Fig. 2h, features both reactive radicals and non-planar $sp^3$ C-H bonds. These characteristics prevent the radicals from strongly bonding to the Au(111) substrate, making them highly reactive for further homocoupling into longer chains. The dashed orange circles in Fig. 2e are CO molecules, which show a preference for adsorption near the terminal units, supporting that the methylene radical-type terminal units have high reactivity.

The methylene radical type terminal units have only one C-H bond dissociated on the zigzag edge. For some terminal units, two C-H bonds would cleavage on one zigzag edge, resulting in azomethine ylide type or diradical type terminal units. The azomethine ylide type also shows different contrasts on the two *ortho* C atoms (Fig. 2f). The *ortho* C atom marked by the red arrow has a brighter contrast, which indicates larger repulsive forces stem from a negative charge on this atom when the unit is in azomethine ylide form. The other *ortho* C atom marked by the yellow arrow, unlike the methylene radical type that shows a darker contrast, has the same contrast as with other C atoms in the conjugated skeleton, which indicates the absence of radical on this *ortho* C atom. Therefore, this *ortho* C atom with normal brightness is the $sp^2$ hybridized C atom in azomethine ylide. The chemical structure of the azomethine ylide type terminal units is depicted in Fig. 2i. For diradical type terminal units, both *ortho* C atoms have a darker hue, as indicated by the yellow arrows in Fig. 2g and its corresponsive chemical structure in Fig. 2j.

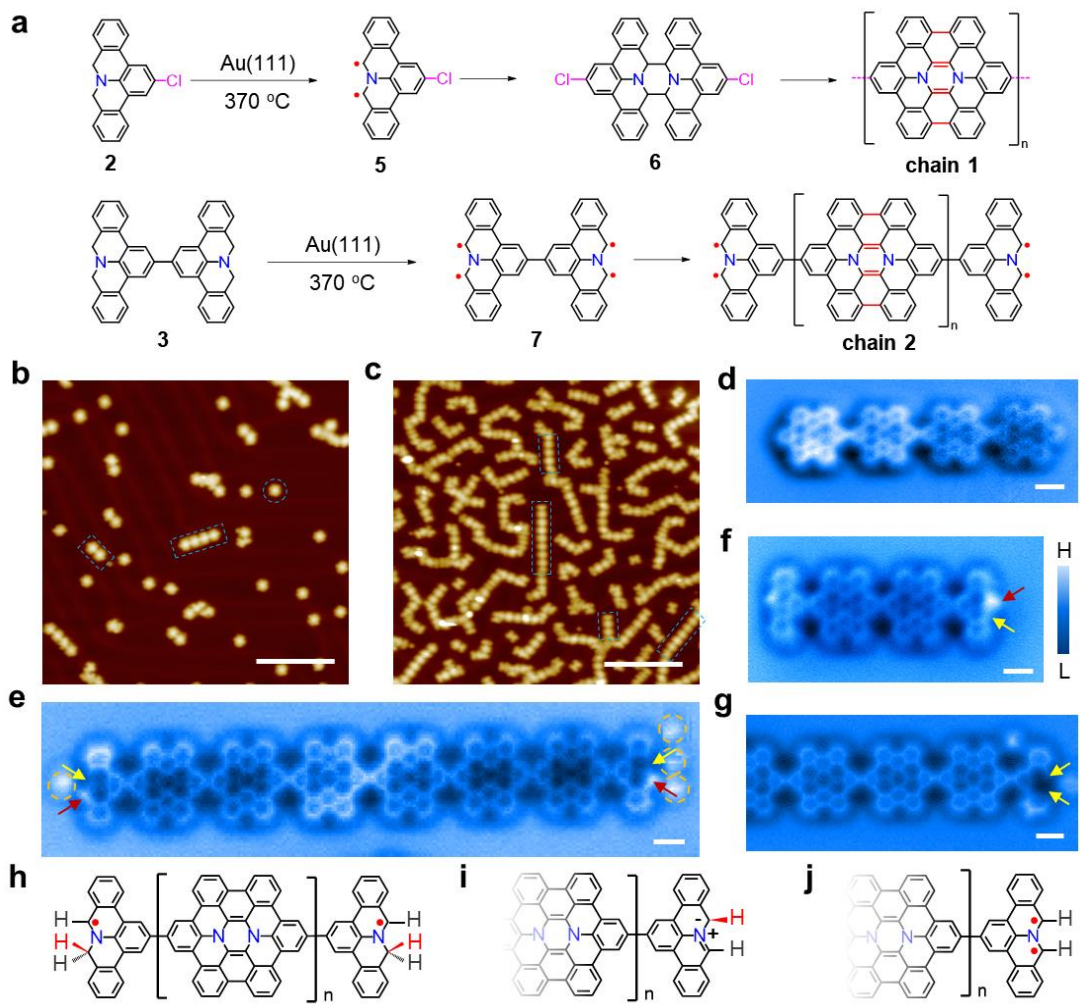

**Fig. 2 On-surface synthesis of diaza-HBC chains.** (**a**) Scheme of the synthesis of diaza-HBC **chain 1** and **chain 2** starting from precursor **2** and **3**. (**b**) Large-are STM image of diaza-HBC **chain 1** after annealing precursor **2** on Au(111) at 370 ℃ (V = -1 V, I= 100 pA). The white dashed boxes denote two **chain 1** with n = 2 and 4, respectively. The white dashed circle marks one of the diaza-HBC monomers. (**c**) Large-area STM image of diaza-HBC **chain 2** after annealing precursor **3** on Au(111) at 370 ℃ (V = -1 V, I= 1 nA). The white dashed boxes denote 4 **chain 2** with n = 2, 3, 5, and 7, respectively. (**d**) Chemical-bond-resolved nc-AFM image of **chain 1** with n = 4 (oscillation amplitude $A_{OSC}$ = 100 pm). (**e**) nc-AFM image of a **chain 2** with n = 6 (oscillation amplitude $A_{OSC}$ = 100 pm). The red arrows highlight the *ortho sp³* C atoms; the yellow arrows denote the radical form of *ortho sp²* C atoms after dehydrogenation. The orange dashed circles mark the CO molecules adsorbed near the terminal units. (**f**) nc-AFM image of a **chain 2** with n = 2, which possesses the azomethine ylide type of terminal units. The red arrow denotes the negatively charged *ortho* C atom; while the yellow arrow marks the other *ortho* C atom with a normal $sp^2$ hybridization. (**g**) nc-AFM image of a **chain 2** exhibiting a diradical type terminal unit. The yellow arrows point to the two *ortho sp²* C atoms, each of which has a radical. (**h**) Corresponsive chemical structure of the chain in (e), highlighting the methylene radical type terminal units. (**i**) Corresponsive chemical structure of the azomethine ylide type terminal units in (f). (**j**) Corresponsive chemical structure of the diradical type terminal units in (g). Scale bars: **b** and **c**, 10 nm, **d - g** 0.5 nm.

On the diaza-HBC chains, we also observed magnetic related electronic states around the Fermi level. Although **chain 1** and **chain 2** have different structures resulting from two different synthetic routes, they exhibit the same prominent odd-even effect, sharing comparable spin excitations on chains with the same unit numbers (Fig. 3, 4 and S10). For even-numbered diaza-HBC chains, a dip at the Fermi level with two symmetric peaks was observed in the dI/dV spectra. For dimers (n = 2 chains), the dI/dV spectra on both units are identical, in which the symmetric peaks are located at ±37 mV, marked by the blue dots in Fig. 3a. This permits us to estimate intermolecular exchange $J$ = 37 meV. The dI/dV map at -37 mV (Fig. 3d) displays that this state has the same distribution pattern on both units. The dimers possess an antiferromagnetic ground state as confirmed by DFT calculations in Fig. S11, suggesting the dip-like feature observed in the dI/dV spectra should be attributed to an inelastic excitation from the ground state to the first excited state of an antiferromagnetic Heisenberg dimer. For n = 4 chains, the dI/dV spectra exhibit inelastic spin excitations with different shapes and intensities between the two outer units (unit 1 and unit 4) and the two inner units (unit 2 and unit 3), as depicted in Fig. 3b. The first and second excitations at ±20 mV and ±50 mV were observed for all the four units, marked by the blue and green dots, respectively. The third excitation at -77 mV can only be observed on the inner units, denoted by orange dots. The dI/dV map taken at the energy of the first excitation (Fig. 3e) shows that the spectra weight is mainly located on the two outer units. For n = 6 chains, the dI/dV spectra exhibit a narrower first excitation gap with the inelastic steps at around ±13 mV (blue dots in Fig. 3c), which has a stronger intensity on units 1, 3, 4, 6 than on units 2 and 5, as shown in Fig. 3f.

A first neighbor Heisenberg $S$ = 1/2 model with antiferromagnetic coupling $J$ was built to simulate the experimental spin excitations. The theoretical dI/dV curves were computed by employing inelastic tunneling theory using perturbation theory up to third order.[45] While second-order perturbation theory gives the inelastic step features, the third-order correction, which accounts for the screening of the substrate, is responsible for the overshooting at the on-set of excitation, as well as the Kondo peak seen in the odd length chains discussed later in the text.[45, 46] Due to spin selection rules, inelastic steps can only appear for transitions to states whose spin differs by at most one unit from that of the ground state. For dimers, the model shows the same dI/dV profile on both units with an excitation step, as denoted by the blue dots in Fig. 3g, which is qualitatively consistent with the experimental observations. The experimental dI/dV is broader than the calculation, because these only include thermal smearing and ignore the broadening due to exchange with the substrate. By increasing the temperature in the simulation, we see that all features become broader, and closer to what is found experimentally (Fig. S13). The step edge energy, which equals to the first excitation energy $J$ of the system, is 37 mV in our case. For n = 4 chains, there are two excitations on all four units, located at $0.659J$ and $1.366J$ according to the theoretical calculations, as marked by the blue and green dots in Fig. 3h. With $J = 37$ mV, the theoretical energy for these two excitations is 24.38 mV and 50.54 mV, which is very close to the first and second spin excitation energies in the experimental data. Furthermore, for the inner units, the theoretical results show an additional third excitation step at $2.073J = 77$ mV, the same as the third excitation observed in Fig. 2b. Therefore, not only the number, but also the energy levels of the spin excitations fit well with the dI/dV spectra, confirming the singlet ground state on the even-numbered diaza-HBC chains. The dI/dV spectra and the theoretically calculated excitations on n = 6 chains are also qualitatively consistent with each other, as presented in Fig. S12a. The calculated spectral weight of the first excitation on n = 2, 4, and 6 chains are shown in Fig. 3i, which accurately reproduces the distribution in the dI/dV maps (Fig. 3d-f). Fig. 3j elucidate that the $S = 1/2$ spin chains have a singlet ground state and a triplet first excited state, where the first excitation energies decrease with longer chains.

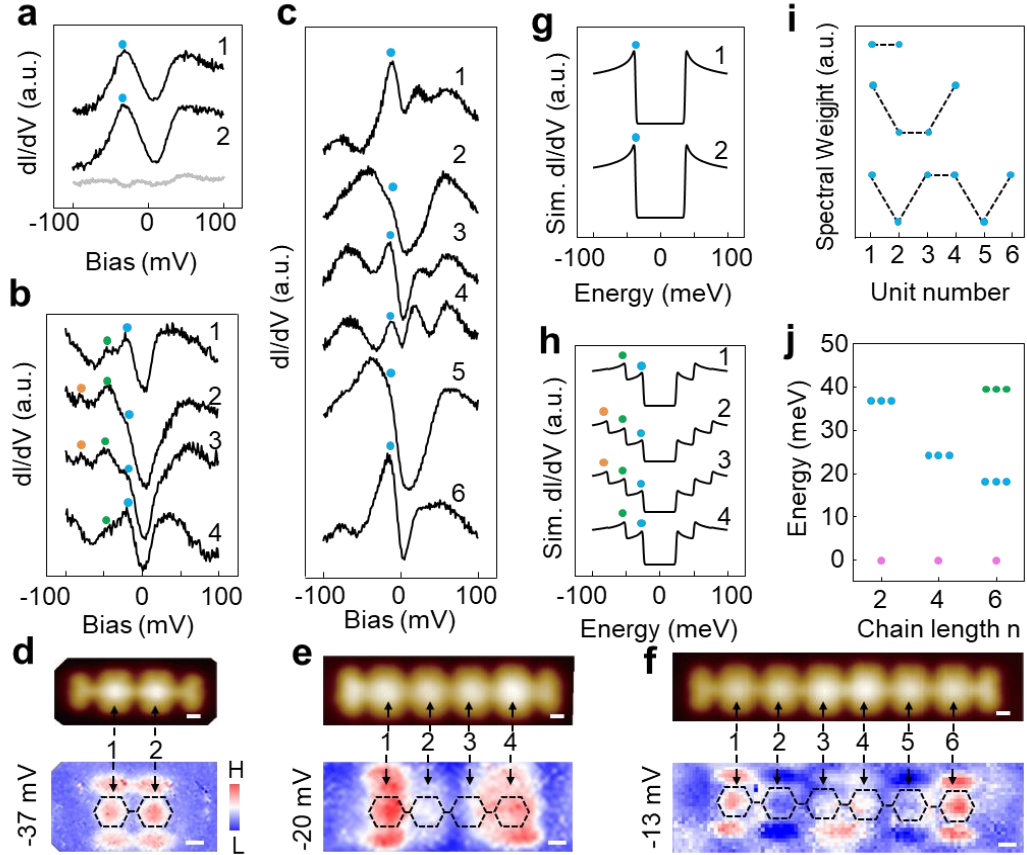

**Fig. 3 Spin excitations in even-numbered diaza-HBC chains.** (**a**) dI/dV spectra taken at the center of the two units in a diaza-HBC dimer ($V_{AC}$ = 1 mV). The excitation edges are denoted by blue dots. The grey curve is the spectrum on Au(111) taken with the same tip on the chains. (**b**) dI/dV spectra taken at the center of the units in an n = 4 diaza-HBC chain ($V_{AC}$ = 1 mV). The first, second, and third spin excitations are denoted by blue, green, and orange dots, respectively. (**c**) dI/dV spectra taken at the center of the units in an n = 6 diaza-HBC chain ($V_{AC}$ = 1 mV). Only the first spin excitations are marked out by blue dots. (**d-f**) STM images (upper panel) of **chain 2** with n = 2, 4, and 6, and the corresponsive dI/dV map (lower panel) at the first spin excitation energies, respectively. (V = -0.1 V, I = 1 nA, $V_{AC}$ = 10 mV). (**g**) and (**h**) Theoretical calculations of spin excitations on n = 2 and 4 diaza-HBC chains, respectively, based on the Heisenberg model $S = 1/2$ chain with an effective temperature of 5 K. The excitation edges are marked by dots with the same color code as in the experimental dI/dV spectra in (a)-(c). (**i**) Theoretical calculated spectral weight of the first excited states of n = 2, 4, and 6 chains. (**j**) Theoretical calculated ground state and the excited states below 50 meV of n = 2, 4, and 6 chains. The pink dots denote the singlet ground state of the chains, while the blue (green) dots denote the triplet's first (second) excited states. Scale bars: 0.5 nm.

For odd-numbered diaza-HBC chains, the dI/dV spectra exhibit a prominent unit-dependent Kondo resonance as shown in Fig. 4. For n = 3 chains (Fig. 4a), the center unit (unit 2) has the same dip-like feature around the Fermi level (blue dots on curve 2), similar to the those on the even-numbered chains. For the two outer units (unit 1 and unit 3), however, an additional peak around the Fermi level appears (marked by the red triangles on curves 1 and 3). For n = 5 chains, the central unit (unit 3) and the two outermost units (unit 1 and unit 5) all present a peak around the Fermi level as indicated by the red triangles on curves 1, 3, and 5 in Fig. 4b; while the two units next to the central unit (unit 2 and unit 4) show similar symmetric peaks as the central unit in n = 3 chains. The n = 7 chains further demonstrate this unit-dependent phenomenon, that the odd-numbered units have an additional zero-bias peak compared to the even-numbered units where only inelastic steps are seen around the Fermi level (Fig. 4c). The dI/dV maps taken at 0 mV in Fig. 4d-f exhibit that the zero-bias peak has large spectral weight on the odd-numbered units with the same spatial distribution pattern as the Kondo peak on the diaza-HBC monomer.

As before, to understand this unit-dependent behavior on the odd-numbered chains, we performed dI/dV simulations using perturbation theory including up to third-order corrections. As shown in Fig. 4g and 4h for chains with units n = 3 and 5, the dI/dV curves on all units show inelastic spin excitations (the first excitations are denoted by blue dots), while only the odd-numbered units have an enhanced Kondo resonance (denoted by red triangles). The calculated inelastic spectra on n = 7 chains are presented in Fig. S12b, which also have enhanced Kondo resonance on the odd-numbered units. The calculated local average magnetization <$S_z$> (Fig. 4i) of the ground state with S=1/2, $S_z = +\frac{1}{2}$ reveals an inhomogeneous distribution across the chain, with antiferromagnetic correlations and larger magnetization, in absolute value, in the odd-numbered sites, that accounts for the larger zero-bias Kondo signal in those sites. Since the odd-numbered chains have a total net spin of $S$ =1/2, the longer the chains the smaller the net spin on each unit. Therefore, the intensity of the Kondo resonance is much weaker in longer chains, as confirmed by the dI/dV map in Fig. 4f, where the Kondo peak on the n = 7 chain is weaker and blurry compared to that on n = 3 and 5 chains in Fig. 4d and 4e. The excitation energies from the doublet ground state to the first excitation doublet state are smaller for longer chains than for shorter chains, as demonstrated in Fig. 4j, consistent with the experimental observations.

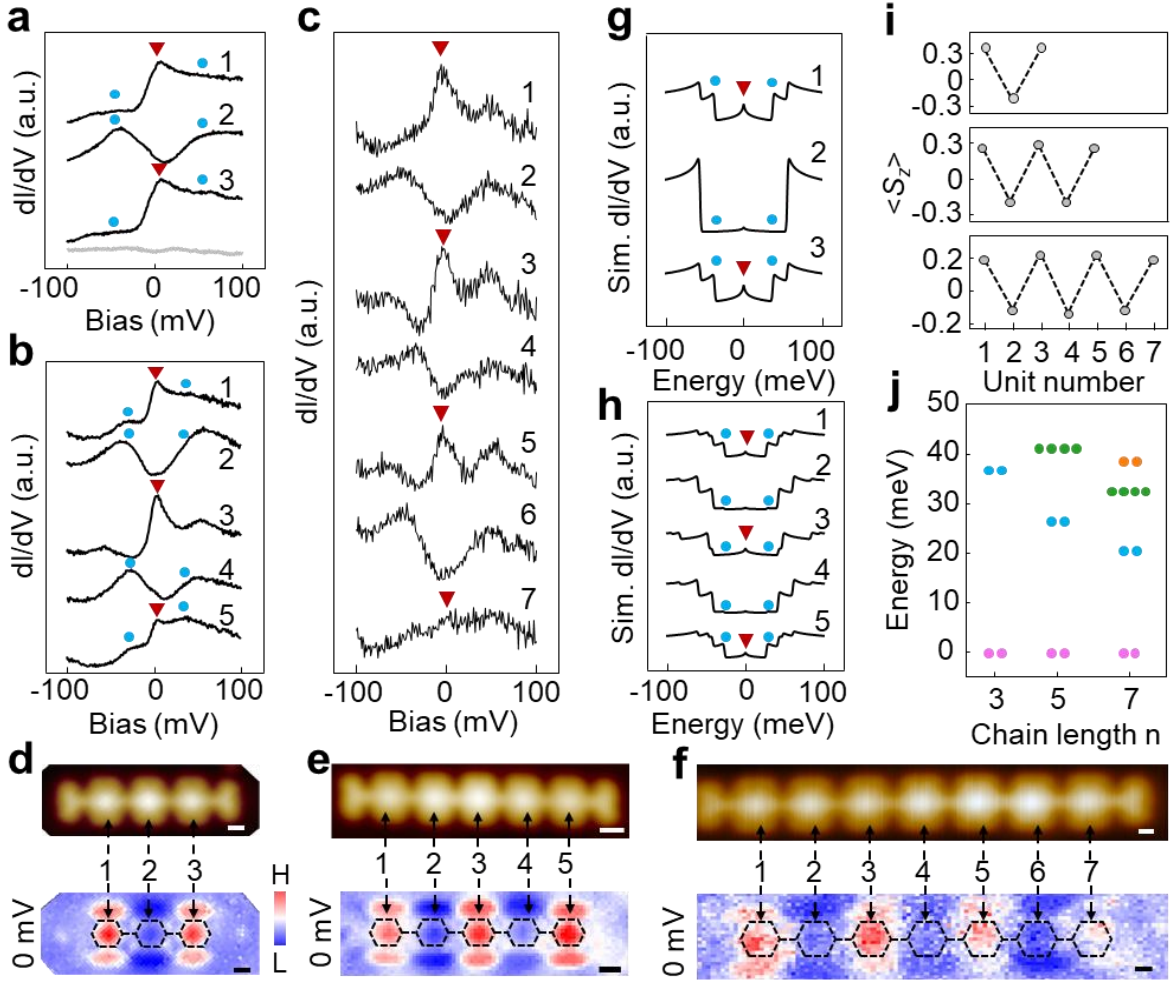

**Fig. 4 Spin excitations in odd-numbered diaza-HBC chains.** (**a**-**c**) dI/dV spectra taken at the center of the units a diaza-HBC chains with n = 3, 5, and 7, respectively ($V_{AC}$ = 1 mV). The red triangles denote the Kondo peak; the blue dots indicate the spin excitation edges. The grey curve is the spectrum on Au(111) taken with the same tip on the chains. (**d**-**f**) STM images (upper panel) of **chain 2** with n = 3, 5, and 6, and the corresponsive dI/dV map (lower panel) at the first spin excitation energies, respectively. (V = -0.1 V, I = 1 nA, $V_{AC}$ = 10 mV). (**g**) and (**h**) Theoretical calculations of dI/dV spectra on n = 3 and 5 diaza-HBC chains, respectively, based on the Heisenberg model $S$ = 1/2 chain with an effective temperature of 5 K. The first spin excitation edges are marked by blue dots; while the Kondo peaks are marked by red triangles. (**i**) Theoretical calculated local average magnetization in $z$ direction $<S_z>$ for the n = 3, 5, and 7 $S$ =1/2 spin chains. (**j**) The theoretically calculated ground state and the excited states below 50 meV of n = 3, 5, and 7 chains. The pink dots denote the doublet ground state of the odd-numbered chains, while the blue, green, and orange dots denote the doublet first excited state, the quartet second excited state, and the doublet third excited state, respectively. Scale bars: 0.5 nm.

## Conclusions

We successfully synthesized spin-1/2 Heisenberg chains based on antiaromatic diaza-HBC units. The chains were constructed through C-H activation of $sp^3$ C atoms at the *ortho* position of a N heteroatom embedded in a zigzag carbon edge. The resultant antiaromatic unit, diaza-HBC monomer, was obtained with a high yield of ~84% on Au(111) substrate. By bond-resolved nc-AFM, we observed three dehydrogenation products at the terminal units of the diaza-HBC chains, which are the methylene radical type, the azomethine ylide type, and the diradical type. The diaza-HBC unit transforms into a radical cation nanomagnet by donating one electron to the Au(111) substrate. The diaza-HBC chains follow a $S = 1/2$ antiferromagnetic Heisenberg chain model, that accounts both for the parity-dependent dI/dV spectra and the inhomogeneous distribution of the zero-bias Kondo peak across different molecules in the odd sites of odd-numbered chains. Our results demonstrated a highly efficient reaction strategy to construct $S = 1/2$ antiferromagnetic spin chains on Au(111) surface, which have potential in organic nano spintronics.[47]

## Methods

**Experimental Methods**

**Sample preparation**: The atomically clean Au(111) crystal surfaces were prepared by several cycles of Ar$^+$ sputtering and annealing in the preparation chamber of an ultra-high vacuum (with a base pressure better than $3.0 \times 10^{-10}$ mbar) low-temperature STM/nc-AFM system (Createc). The molecules were deposited onto the clean surface and verified by STM. The annealing was realized by a direct filament heating.

**STM/AFM**: The STM and nc-AFM measurements were performed at 4.5 K in UHV conditions. A qPlus sensor[48] consisting of a Pt/Ir tip with a resonance frequency of 28.9 kHz was used for both STM and AFM measurements. For STM, the tip is holding at feedback tunneling current ($I_t$) and a sample bias ($V_s$), and all STM images were taken with constant current mode. For nc-AFM, the tip was firstly stabilized at a tunneling junction of a current ($I_t$) and a sample bias ($V_s$), and then worked in constant height mode with $V_s = 0$ V. The bond-resolution nc-AFM images were taken with a CO functionalized tip[49, 50] in constant-height mode with an oscillation amplitude of 100 pm.

**Theoretical Modelling**

We used a Heisenberg model for spin chains with n units, $\sum_{i=1}^{i=N-1} \vec{S}_i \cdot \vec{S}_{i+1}$, with $J = 37$ meV. The eigenstates of this Heisenberg Hamiltonian (HH) have a well-defined total spin $S$, and third component $S_z$. For the calculation of the dI/dV we use single-spin inelastic electron tunnel theory,[46] up to third-order[45] in the exchange interaction between the surface spin and the tip and substrate electrons. In this approach, inelastic features are predicted when the eV matches the energy difference between eigenstates of the HH whose spin $S$ differs at most in one unit. The height of the steps associated to a given transition is controlled by the spin spectral weight, defined as $|\langle GS|S^a(i)|X\rangle|^2$, where GS and X refer to the wave functions of the ground state and excited state, respectively, and $S^a(i)$ is the $a = x, y, z$ component of the spin operator at site $i$. The broadening of the steps reflects thermal smearing of the electrons in tip and substrate. Other sources of broadening, such as finite spin-lifetime of the HH states due to Kondo coupling to the substrate, are not taken into account.

**DFT calculations**

All DFT calculations were carried out using the Vienna ab-initio simulation package (VASP)[51] with the projector augmented wave (PAW) method.[52, 53] The van der Waals (vdW) interactions were considered at the vdW-DF level,[54] with the optB86 functional used for the exchange potential.[55, 56] The energy cutoff of the plane-wave basis sets was 400 eV. The vacuum layer was larger than 1.2 nm. The K-point sampling was done only at the Γ point. For geometry optimization, the bottom one Au layer was kept fixed while the molecules and the top two layers Au were allowed to relax until atomic forces became lower than 0.1 eV/nm.

## Data availability

The authors declare that the data supporting the findings of this study are available within the paper and its Supplementary Information files. The data that support the findings of this study are available from the corresponding authors upon request.

## Acknowledgements

We thank for the helpful discussions with Prof. Ziqiang Wang and Prof. Feng Liu. We acknowledge the financial support from the National Natural Science Foundation of China (No. 62488201), the National Key Research and Development Program of China (No. 2022YFA1204100), the German Research Foundation (DFG) with EnhanceNano (No. 391979941), the EU Graphene Flagship (Graphene Core 3, 881603), ERC Consolidator Grant (T2DCP, 819698), the Center for Advancing Electronics Dresden (cfaed), H2020-EU.1.2.2.—FET Proactive Grant (LIGHT-CAP, 101017821), the CAS Project for Young Scientists in Basic Research (YSBR-003, YSBR-053), the Swiss SNF Pimag Grant, the HE Grant FUNLAYERS- 101079184, and the Innovation Program of Quantum Science and Technology (2021ZD0302700).

## Author Contributions

H.-J.G. and X.F. supervised the project; L.H., J.M. and. C.-A.P. conceived the experiment; X.S.F., L.H., X.H., H.C., Z.C. and X.L. performed STM/AFM experiments with the guidance of H.-J.G.; K.L. synthesized and characterized the precursor molecules under the supervision of J.M. and X.F.; J.C.G.H. performed theoretical calculations supervised by J.F.R.; Y.G. and S.D. carried out DFT calculations. L.H., J.M., J.F.R, X.F. and H.-J.G. wrote the manuscript with inputs from all authors. X.S.F., L.H., and K.L. contributed equally to this work.

**References**

1. J. Fernández-Rossier, J. J. Palacios, Magnetism in Graphene Nanoislands. *Phys. Rev. Lett.* **99**, 177204 (2007).
2. W. Han, R. K. Kawakami, M. Gmitra, J. Fabian, Graphene spintronics. *Nat. Nanotech.* **9**, 794-807 (2014).
3. O. V. Yazyev, M. I. Katsnelson, Magnetic Correlations at Graphene Edges: Basis for Novel Spintronics Devices. *Phys. Rev. Lett.* **100**, 047209 (2008).
4. H. Wang, H. S. Wang, C. Ma, L. Chen, C. Jiang, C. Chen, X. Xie, A.-P. Li, X. Wang, Graphene nanoribbons for quantum electronics. *Nat. Rev. Phys.* **3**, 791-802 (2021).
5. A. Candini, S. Klyatskaya, M. Ruben, W. Wernsdorfer, M. Affronte, Graphene Spintronic Devices with Molecular Nanomagnets. *Nano Lett.* **11**, 2634-2639 (2011).
6. A. R. Rocha, V. M. García-suárez, S. W. Bailey, C. J. Lambert, J. Ferrer, S. Sanvito, Towards molecular spintronics. *Nat. Mater.* **4**, 335-339 (2005).
7. D. Pesin, A. H. MacDonald, Spintronics and pseudospintronics in graphene and topological insulators. *Nat. Mater.* **11**, 409-416 (2012).
8. Y. Zheng, C. Li, Y. Zhao, D. Beyer, G. Wang, C. Xu, X. Yue, Y. Chen, D.-D. Guan, Y.-Y. Li, H. Zheng, C. Liu, W. Luo, X. Feng, S. Wang, J. Jia, Engineering of Magnetic Coupling in Nanographene. *Phys. Rev. Lett.* **124**, 147206 (2020).
9. Y. Zheng, C. Li, C. Xu, D. Beyer, X. Yue, Y. Zhao, G. Wang, D. Guan, Y. Li, H. Zheng, C. Liu, J. Liu, X. Wang, W. Luo, X. Feng, S. Wang, J. Jia, Designer spin order in diradical nanographenes. *Nat. Commun.* **11**, 6076 (2020).
10. J. Li, S. Sanz, M. Corso, D. J. Choi, D. Peña, T. Frederiksen, J. I. Pascual, Single spin localization and manipulation in graphene open-shell nanostructures. *Nat. Commun.* **10**, 200 (2019).
11. S. Mishra, D. Beyer, K. Eimre, S. Kezilebieke, R. Berger, O. Gröning, C. A. Pignedoli, K. Müllen, P. Liljeroth, P. Ruffieux, X. Feng, R. Fasel, Topological frustration induces unconventional magnetism in a nanographene. *Nat. Nanotech.* **15**, 22-28 (2020).
12. J. Li, S. Sanz, J. Castro-Esteban, M. Vilas-Varela, N. Friedrich, T. Frederiksen, D. Pena, J. I. Pascual, Uncovering the Triplet Ground State of Triangular Graphene Nanoflakes Engineered with Atomic Precision on a Metal Surface. *Phys. Rev. Lett.* **124**, 177201 (2020).
13. P. Ruffieux, S. Wang, B. Yang, C. Sánchez-Sánchez, J. Liu, T. Dienel, L. Talirz, P. Shinde, C. A. Pignedoli, D. Passerone, T. Dumslaff, X. Feng, K. Müllen, R. Fasel, On-surface synthesis of graphene nanoribbons with zigzag edge topology. *Nature* **531**, 489-492 (2016).
14. S. Mishra, X. Yao, Q. Chen, K. Eimre, O. Groening, R. Ortiz, M. Di Giovannantonio, J. C. Sancho-Garcia, J. Fernandez-Rossier, C. A. Pignedoli, K. Muellen, P. Ruffieux, A. Narita, R. Fasel, Large magnetic exchange coupling in rhombus-shaped nanographenes with zigzag periphery. *Nat. Chem.* **13**, 581-586 (2021).
15. S. Mishra, D. Beyer, K. Eimre, R. Ortiz, J. Fernandez-Rossier, R. Berger, O. Groening, C. A. Pignedoli, R. Fasel, X. Feng, P. Ruffieux, Collective All-Carbon Magnetism in Triangulene Dimers. *Angew. Chem. Int. Ed.* **59**, 12041-12047 (2020).
16. N. Pavlíček, A. Mistry, Z. Majzik, N. Moll, G. Meyer, D. J. Fox, L. Gross, Synthesis and characterization of triangulene. *Nat. Nanotech.* **12**, 308-311 (2017).
17. R. E. Blackwell, F. Zhao, E. Brooks, J. Zhu, I. Piskun, S. Wang, A. Delgado, Y.-L. Lee, S. G. Louie, F. R. Fischer, Spin splitting of dopant edge state in magnetic zigzag graphene nanoribbons. *Nature* **600**,

647-652 (2021).

18. Q. Du, X. Su, Y. Liu, Y. Jiang, C. Li, K. Yan, R. Ortiz, T. Frederiksen, S. Wang, P. Yu, Orbital-symmetry effects on magnetic exchange in open-shell nanographenes. *Nat. Commun.* **14**, 4802 (2023).

19. S. Song, A. Pinar Solé, A. Matěj, G. Li, O. Stetsovych, D. Soler, H. Yang, M. Telychko, J. Li, M. Kumar, Q. Chen, S. Edalatmanesh, J. Brabec, L. Veis, J. Wu, P. Jelinek, J. Lu, Highly entangled polyradical nanographene with coexisting strong correlation and topological frustration. *Nat. Chem.* **16**, 938-944 (2024).

20. X. Chang, L. Huang, Y. Gao, Y. Fu, J. Ma, H. Yang, J. Liu, X. Fu, X. Lin, X. Feng, S. Du, H.-J. Gao, On-surface synthesis and edge states of NBN-doped zigzag graphene nanoribbons. *Nano Res.* **16**, 10436-10442 (2023).

21. S. Mishra, D. Beyer, K. Eimre, J. Liu, R. Berger, O. Gröning, C. A. Pignedoli, K. Müllen, R. Fasel, X. Feng, P. Ruffieux, Synthesis and Characterization of π-Extended Triangulene. *J. Am. Chem. Soc.* **141**, 10621-10625 (2019).

22. Q. Sun, X. Yao, O. Gröning, K. Eimre, C. A. Pignedoli, K. Müllen, A. Narita, R. Fasel, P. Ruffieux, Coupled Spin States in Armchair Graphene Nanoribbons with Asymmetric Zigzag Edge Extensions. *Nano Lett.* **20**, 6429-6436 (2020).

23. R. D. McCurdy, A. Delgado, J. Jiang, J. Zhu, E. C. H. Wen, R. E. Blackwell, G. C. Veber, S. Wang, S. G. Louie, F. R. Fischer, Engineering Robust Metallic Zero-Mode States in Olympicene Graphene Nanoribbons. *J. Am. Chem. Soc.* **145**, 15162-15170 (2023).

24. O. V. Yazyev, Emergence of magnetism in graphene materials and nanostructures. *Rep. Prog. Phys.* **73**, 056501 (2010).

25. F. D. M. Haldane, Continuum dynamics of the 1-D Heisenberg antiferromagnet: Identification with the O(3) nonlinear sigma model. *Phys. Lett. A* **93**, 464-468 (1983).

26. F. D. M. Haldane, Nonlinear Field Theory of Large-Spin Heisenberg Antiferromagnets: Semiclassically Quantized Solitons of the One-Dimensional Easy-Axis Néel State. *Phys. Rev. Lett.* **50**, 1153-1156 (1983).

27. M. Hagiwara, K. Katsumata, I. Affleck, B. I. Halperin, J. P. Renard, Observation of S=1/2 degrees of freedom in an S=1 linear-chain Heisenberg antiferromagnet. *Phys. Rev. Lett.* **65**, 3181-3184 (1990).

28. S. Miyashita, S. Yamamoto, Effects of edges in S=1 Heisenberg antiferromagnetic chains. *Phys. Rev. B* **48**, 913-919 (1993).

29. S. Lounis, P. H. Dederichs, S. Blügel, Magnetism of Nanowires Driven by Novel Even-Odd Effects. *Phys. Rev. Lett.* **101**, 107204 (2008).

30. A. Machens, N. P. Konstantinidis, O. Waldmann, I. Schneider, S. Eggert, Even-odd effect in short antiferromagnetic Heisenberg chains. *Phys. Rev. B* **87**, 144409 (2013).

31. P. Politi, M. G. Pini, Even-odd effects in finite Heisenberg spin chains. *Phys. Rev. B* **79**, 012405 (2009).

32. S. Mishra, G. Catarina, F. Wu, R. Ortiz, D. Jacob, K. Eimre, J. Ma, C. A. Pignedoli, X. Feng, P. Ruffieux, J. Fernández-Rossier, R. Fasel, Observation of fractional edge excitations in nanographene spin chains. *Nature* **598**, 287-292 (2021).

33. Y. Zhao, K. Jiang, C. Li, Y. Liu, G. Zhu, M. Pizzochero, E. Kaxiras, D. Guan, Y. Li, H. Zheng, C. Liu, J. Jia, M. Qin, X. Zhuang, S. Wang, Quantum nanomagnets in on-surface metal-free porphyrin chains. *Nat. Chem.* **15**, 53-60 (2023).

34. A. S. da Costa Azevêdo, A. Saraiva-Souza, V. Meunier, E. C. Girão, Electronic properties of N-rich graphene nano-chevrons. *Phys. Chem. Chem. Phys.* **23**, 13204-13215 (2021).

35. K. Sun, O. J. Silveira, S. Saito, K. Sagisaka, S. Yamaguchi, A. S. Foster, S. Kawai, Manipulation of

Spin Polarization in Boron-Substituted Graphene Nanoribbons. *ACS Nano* **16**, 11244-11250 (2022).

36. E. C. H. Wen, P. H. Jacobse, J. Jiang, Z. Wang, R. D. McCurdy, S. G. Louie, M. F. Crommie, F. R. Fischer, Magnetic Interactions in Substitutional Core-Doped Graphene Nanoribbons. *J. Am. Chem. Soc.* **144**, 13696-13703 (2022).

37. E. C. H. Wen, P. H. Jacobse, J. Jiang, Z. Wang, S. G. Louie, M. F. Crommie, F. R. Fischer, Fermi-Level Engineering of Nitrogen Core-Doped Armchair Graphene Nanoribbons. *J. Am. Chem. Soc.* **145**, 19338-19346 (2023).

38. N. Friedrich, P. Brandimarte, J. Li, S. Saito, S. Yamaguchi, I. Pozo, D. Peña, T. Frederiksen, A. Garcia-Lekue, D. Sánchez-Portal, J. I. Pascual, Magnetism of Topological Boundary States Induced by Boron Substitution in Graphene Nanoribbons. *Phys. Rev. Lett.* **125**, 146801 (2020).

39. X. Y. Wang, M. Richter, Y. He, J. Bjork, A. Riss, R. Rajesh, M. Garnica, F. Hennersdorf, J. J. Weigand, A. Narita, R. Berger, X. Feng, W. Auwarter, J. V. Barth, C. A. Palma, K. Mullen, Exploration of pyrazine-embedded antiaromatic polycyclic hydrocarbons generated by solution and on-surface azomethine ylide homocoupling. *Nat. Commun.* **8**, 1948 (2017).

40. R. Berger, M. Wagner, X. Feng, K. Müllen, Polycyclic aromatic azomethine ylides: a unique entry to extended polycyclic heteroaromatics. *Chem. Sci.* **6**, 436-441 (2015).

41. Q. Q. Li, K. Ochiai, C. A. Lee, S. Ito, Synthesis of pi-Extended Imidazoles by 1,3-Dipolar Cycloaddition of Polycyclic Aromatic Azomethine Ylides with Nitriles. *Org. Lett.* **22**, 6132-6137 (2020).

42. A. Riss, M. Richter, A. P. Paz, X. Y. Wang, R. Raju, Y. He, J. Ducke, E. Corral, M. Wuttke, K. Seufert, M. Garnica, A. Rubio, V. B. J, A. Narita, K. Mullen, R. Berger, X. Feng, C. A. Palma, W. Auwarter, Polycyclic aromatic chains on metals and insulating layers by repetitive [3+2] cycloadditions. *Nat. Commun.* **11**, 1490 (2020).

43. Y. Gao, L. Huang, Y. Cao, M. Richter, J. Qi, Q. Zheng, H. Yang, J. Ma, X. Chang, X. Fu, C.-A. Palma, H. Lu, Y.-Y. Zhang, Z. Cheng, X. Lin, M. Ouyang, X. Feng, S. Du, H.-J. Gao, Selective activation of four quasi-equivalent C–H bonds yields N-doped graphene nanoribbons with partial corannulene motifs. *Nat. Commun.* **13**, 6146 (2022).

44. S. Kawai, S. Nakatsuka, T. Hatakeyama, R. Pawlak, T. Meier, J. Tracey, E. Meyer, A. S. Foster, Multiple heteroatom substitution to graphene nanoribbon. *Sci. Adv.* **4**, eaar7181 (2018).

45. M. Ternes, Spin excitations and correlations in scanning tunneling spectroscopy. *New J. Phys.* **17**, 063016 (2015).

46. J. Fernández-Rossier, Theory of Single-Spin Inelastic Tunneling Spectroscopy. *Phys. Rev. Lett.* **102**, 256802 (2009).

47. During the final stage of completion of this work, we became aware of a similar work by K. Sun *et al.* (arXiv 2407.02142v1, posted on July 2). .

48. F. J. Giessibl, The qPlus sensor, a powerful core for the atomic force microscope. *Rev. Sci. Instrum.* **90**, 011101 (2019).

49. L. Gross, F. Mohn, N. Moll, P. Liljeroth, G. Meyer, The Chemical Structure of a Molecule Resolved by Atomic Force Microscopy. *Science* **325**, 1110-1114 (2009).

50. L. Bartels, G. Meyer, K.-H. Rieder, Controlled vertical manipulation of single CO molecules with the scanning tunneling microscope: A route to chemical contrast. *Appl. Phys. Lett.* **71**, 213-215 (1997).

51. G. Kresse, J. Furthmüller, Efficient iterative schemes for ab initio total-energy calculations using a plane-wave basis set. *Phys. Rev. B* **54**, 11169-11186 (1996).

52. P. E. Blochl, Projector augmented-wave method. *Phys. Rev. B* **50**, 17953-17979 (1994).

53. G. Kresse, D. Joubert, From ultrasoft pseudopotentials to the projector augmented-wave method. *Phys.*

*Rev. B* **59**, 1758-1775 (1999).
54. M. Dion, H. Rydberg, E. Schröder, D. C. Langreth, B. I. Lundqvist, Van der Waals Density Functional for General Geometries. *Phys. Rev. Lett.* **92**, 246401 (2004).
55. J. Klimeš, D. R. Bowler, A. Michaelides, Chemical accuracy for the van der Waals density functional. *J. Phys. Condens. Matter* **22**, 022201 (2009).
56. J. Klimeš, D. R. Bowler, A. Michaelides, Van der Waals density functionals applied to solids. *Phys. Rev. B* **83**, 195131 (2011).

Supplementary Information for

# Building spin-1/2 antiferromagnetic Heisenberg chains with diaza-nanographenes

Xiaoshuai Fu[1,2†], Li Huang[1,2,7†*], Kun Liu[3†], João C. G. Henriques[5,9], Yixuan Gao[1,2], Xianghe Han[1,2], Hui Chen[1,2], Yan Wang[1,2], Carlos-Andres Palma[1,2], Zhihai Cheng[6], Xiao Lin[2,1], Shixuan Du[1,2], Ji Ma[4,8*], Joaquín Fernández-Rossier[5*], Xinliang Feng[3,8], Hong-Jun Gao[1,2,7]

[1] *Beijing National Center for Condensed Matter Physics and Institute of Physics, Chinese Academy of Sciences, Beijing 100190, China.*

[2] *School of Physical Sciences, University of Chinese Academy of Sciences, Beijing 100190, China*

[3] *Center for Advancing Electronics Dresden (cfaed) & Faculty of Chemistry and Food Chemistry, Technische Universität Dresden, D-01069 Dresden, Germany*

[4] *College of Materials Science and Optoelectronic Technology, University of Chinese Academy of Sciences, Beijing 100190, China*

[5] *International Iberian Nanotechnology Laboratory, 4715-310 Braga, Portugal*

[6] *Beijing Key Laboratory of Optoelectronic Functional Materials & Micro-Nano Devices, Department of Physics, Renmin University of China, Beijing 100872, China*

[7] *Hefei National Laboratory, Hefei, Anhui 230088, China*

[8] *Max Planck Institute of Microstructure Physics, Weinberg 2, 06120 Halle, Germany*

[9] *Universidade de Santiago de Compostela, 15782 Santiago de Compostela, Spain*

\* Corresponding author. Email: lhuang@iphy.ac.cn; maji@ucas.ac.cn; joaquin.fernandez-rossier@inl.int

† These authors contributed equally to this work.

## Synthesis of Molecular Precursors

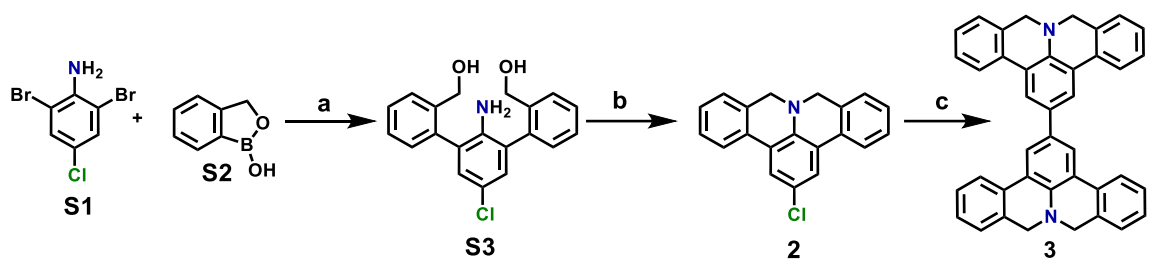

**Scheme S1**. Synthetic route toward monomeric precursor**s** **2** and **3**, as well as their corresponding spin chains. Reagents and conditions: (a) Pd(PPh$_3$)$_4$, K$_2$CO$_3$, toluene/EtOH/H$_2$O, 110 °C, 24 h, 61%; (b) HCl in dioxane (4 M), microwave, 130 °C, 90 min, 72%; (c) Ni(COD)$_2$, COD, 2,2′-bipyridine, toluene/DMF, 80 °C, 2 days, 68%.

The synthesis of the molecular precursors **2** and **3** was depicted in Scheme 1. First, 2,6-diaryl-4-chloroaniline (**S3**) was prepared via 2-fold Suzuki coupling of 2,6-dibromo-4-chloroaniline (**S1**) and 1-hydroxy-3*H*-2,1-benzoxaborole (**S2**) in 61% yield. Afterward, the HCl-catalyzed cyclization of **S3** under microwave conditions yielded 2-chloro-8*H*,10*H*-isoquinolino[4,3,2-*de*]phenanthridine molecule (**2**) as a yellow-green solid in 72% yield. Finally, a nickel-mediated Yamamoto dimerization of compound **2** afforded the biisoquinolino[4,3,2-*de*]phenanthridine precursor (**3**) with a yield of 68%.

## Surface-Assisted Growth of Spin Chain 1 and Chain 2

The monomer precursor **2** was sublimated at 85°C onto Au (111) surface held at 370°C under ultrahigh vacuum (UHV) and detected by scanning tunneling microscopy (STM) (Figure 2 and S8). Dehydrogenation of precursor **2** transforms it into diradical intermediate (**5**) or azomethine ylide form, which further undergoes homocoupling into **6**. Subsequent dechlorination, C-C coupling, and cyclodehydrogenation occur to yield **chain 1**. The monomer precursor **3** was sublimated onto a Au(111) surface under ultrahigh vacuum (UHV) at 370 °C and detected by scanning tunneling microscopy (STM) (Figure 2 and S9). Furthermore, thermally-induced deprotonation of **3** by annealing at 370 °C generated an azomethine ylide (AMY) intermediate (**7**), which exhibits zwitterionic and diradical resonance structures on each side. At the employed temperature, the intermediate underwent azomethine ylide homocoupling and surface-assisted cyclodehydrogenation synergistically, facilitating a step-growth polymerization and forming diaza-HBC-based polymer **chain 2** with different lengths.

## (2'-amino-5'-chloro-[1,1':3',1''-terphenyl]-2,2''-diyl)dimethanol (S3)

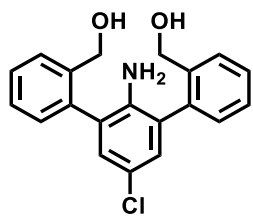

A mixture of compound **S1** (3.71 g, 13.0 mmol), compound **S2** (5.22 g, 39.0 mmol) and K$_2$CO$_3$ (33.24 g, 240.5 mmol) in toluene (200 mL), EtOH (40 mL) and water (40 mL) was degassed for 30 min. Subsequently, Pd(PPh$_3$)$_4$ (750.0 mg, 0.65 mmol) was added. The reaction mixture was refluxed at 110 °C for 24 h under argon. After cooling to room temperature, the mixture was extracted three times with ethyl acetate, washed with brine and dried over MgSO$_4$. Afterward, the solvent was removed under vacuum and the residue was purified by silica gel column chromatography with *iso*-hexane: ethyl acetate (3:2) as eluent to give the compound **S3** as a pale grey solid (2.69 g, 61%). $^1$H NMR (300 MHz, DMSO-d$_6$): 7.63–7.61 (m, 2H), 7.46–7.41 (m, 2H), 7.37–7.32 (m, 2H), 7.20–7.16 (m, 2H), 6.96–6.95 (m, 2H), 5.13–5.05 (m, 2H), 4.43–4.26 (m, 4H), 3.76–3.75 (m, 2H). $^{13}$C NMR (75 MHz, DMSO-d$_6$): 140.97, 140.93, 140.79, 140.55, 135.51, 135.32, 129.64, 129.49, 128.84, 128.69, 128.44, 128.40, 128.03, 128.01, 127.19, 127.17, 127.09, 127.03, 119.76, 119.67, 60.44, 60.24.

**2-chloro-8*H*,10*H*-isoquinolino[4,3,2-*de*]phenanthridine (2)**

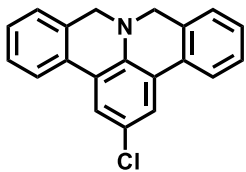

In a dry microwave tube, compound **S3** (200 mg, 588.5 mmol) was added in anhydrous hydrogen chloride solution (4.0 M in dioxane, 2 mL, 8 mmol) under inert conditions. The reaction mixture was placed in a microwave reactor. A dynamic mode was chosen (300 W, power max: on, activated cooling, pre-stirring: 10s, temperature: 130 °C) for 90 min. After cooling to room temperature, the reaction mixture was transferred to the glovebox. The precipitate was filtered and washed with dry MeOH, giving compound **2** a yellow-green solid (128.7 mg, 72%). $^1$H NMR (300 MHz, CD$_2$Cl$_2$): 7.64–7.61 (m, 2H), 7.61 (s, 2H), 7.38–7.32 (m, 2H), 7.31–7.26 (m, 2H), 7.22–7.19 (m, 2H), 4.24 (s, 4H). $^{13}$C NMR (75 MHz, CD$_2$Cl$_2$): 143.24, 131.71, 131.10, 128.49, 128.41, 126.47, 124.56, 123.93, 123.01, 122.84, 54.14. HRMS (MALDI-TOF, m/z): calcd for [C$_{20}$H$_{13}$ClN]$^+$ [M-H]$^+$, 302.0732; observed 302.0313.

**8*H*,8'*H*,10*H*,10'*H*-2,2'-biisoquinolino[4,3,2-*de*]phenanthridine (3)**

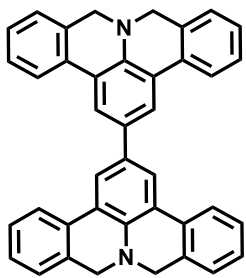

A mixture of compound **2** (50 mg, 164.6 μmol), 2,2'-bipyridine (102.83 mg, 658.3 μmol), bis(cyclooctadiene)nickel(0) (170.47 mg, 658.3 μmol) and cyclooctadiene (71.22 mg, 658.3 μmol) were dissolved in a mixed solvent of dimethyl formamide (DMF) (1 mL) and toluene (4 mL). The reaction mixture was stirred at 80 °C for 2 days in the dark. After cooling to room temperature, the solvent was evaporated under reduced pressure and then transferred to the glovebox. After precipitation from DCM/*n*-pentane, compound **3** was obtained as a pale-yellow solid (30 mg, 68%). $^1$H NMR (300 MHz, CD$_2$Cl$_2$): 7.99 (s, 2H), 7.86–7.83 (m, 2H), 7.42–7.37 (m, 2H), 7.33–7.28 (m, 2H), 7.26–7.24 (m, 2H), 4.30 (s, 4H). $^{13}$C NMR (75 MHz, CD$_2$Cl$_2$): 143.77, 132.52, 132.20, 131.86, 128.39, 127.92, 126.46, 122.91, 122.76, 121.78, 54.41. HRMS (MALDI-TOF, m/z): calcd for [C$_{40}$H$_{27}$N$_2$]$^+$ [M-H]$^+$, 535.2169; observed 535.2180.

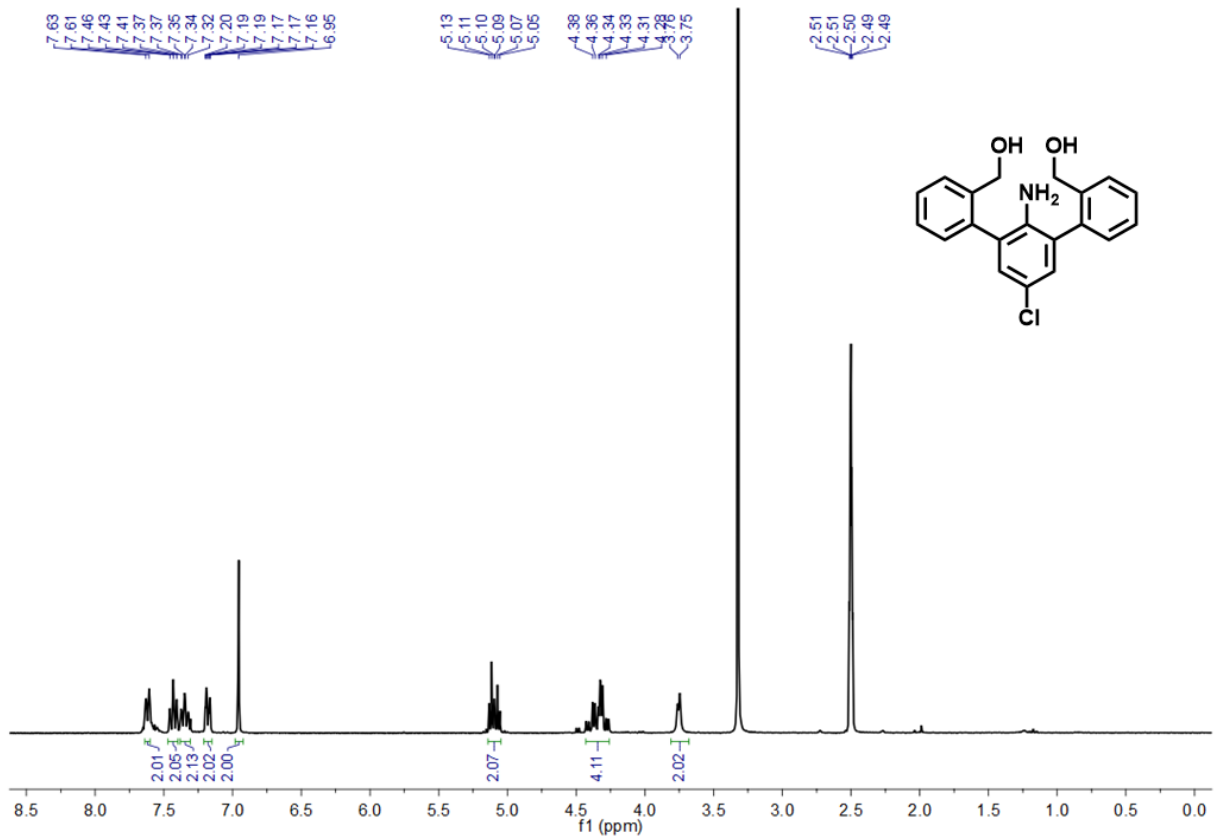

**Figure S1.** $^1$H NMR spectrum (300 MHz) of **S3** in DMSO-d$_6$ at room temperature.

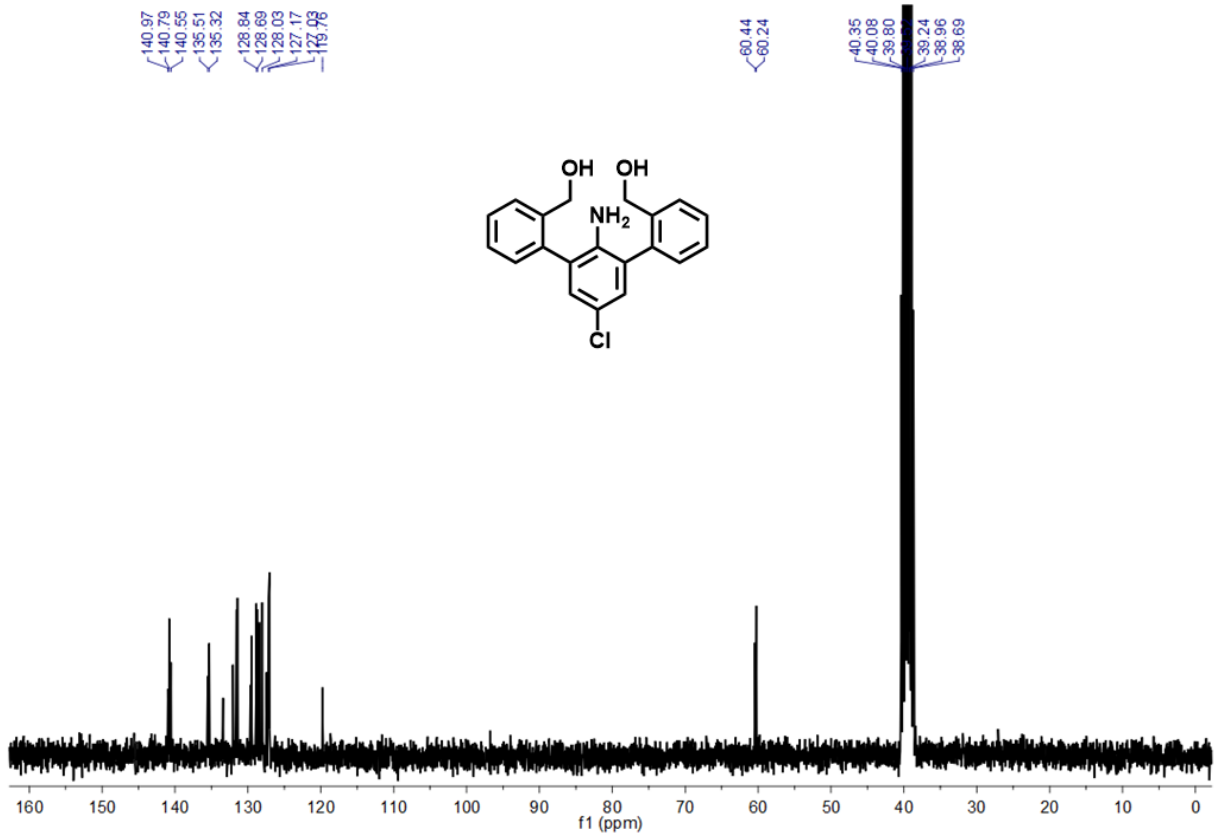

**Figure S2.** $^{13}$C NMR spectrum (75 MHz) of **S3** in DMSO-d$_6$ at room temperature.

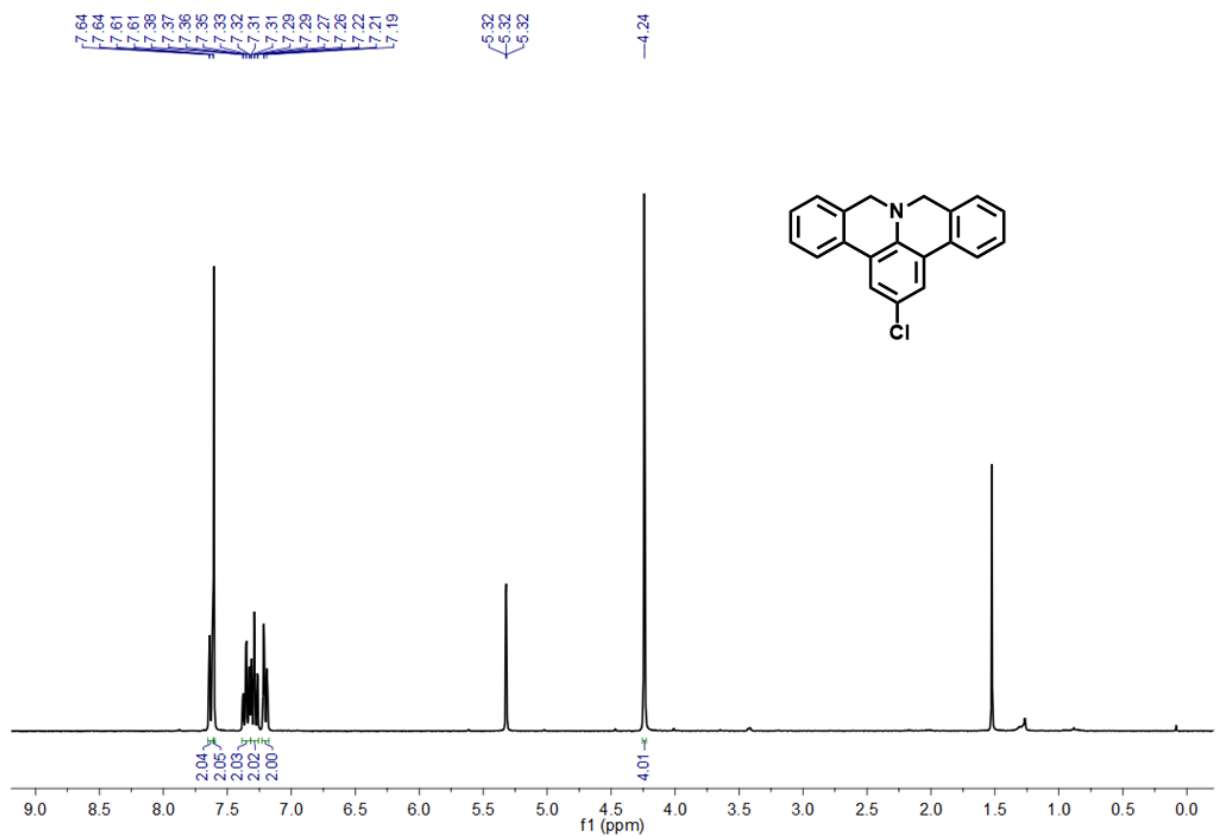

**Figure S3.** [1]H NMR spectrum (300 MHz) of **2** in $CD_2Cl_2$ at room temperature.

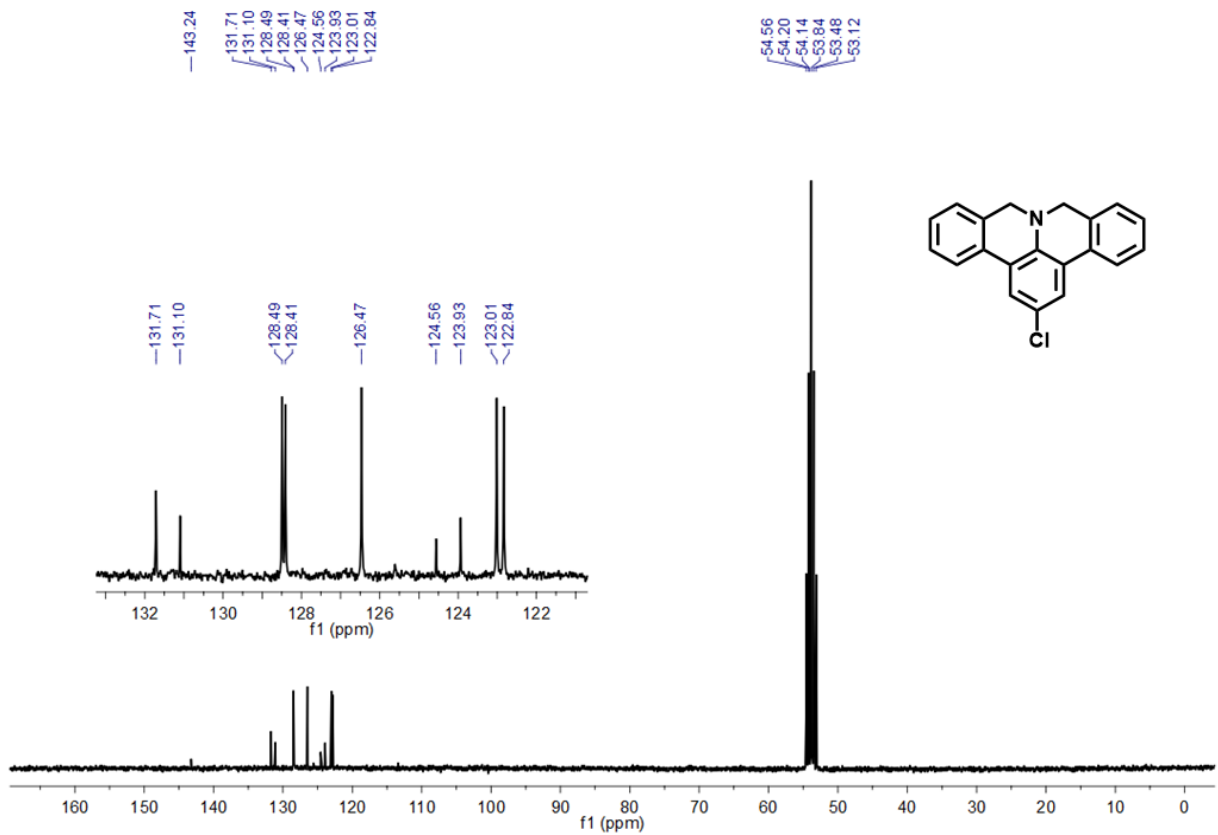

**Figure S4.** [13]C NMR spectrum (75 MHz) of **2** in $CD_2Cl_2$ at room temperature.

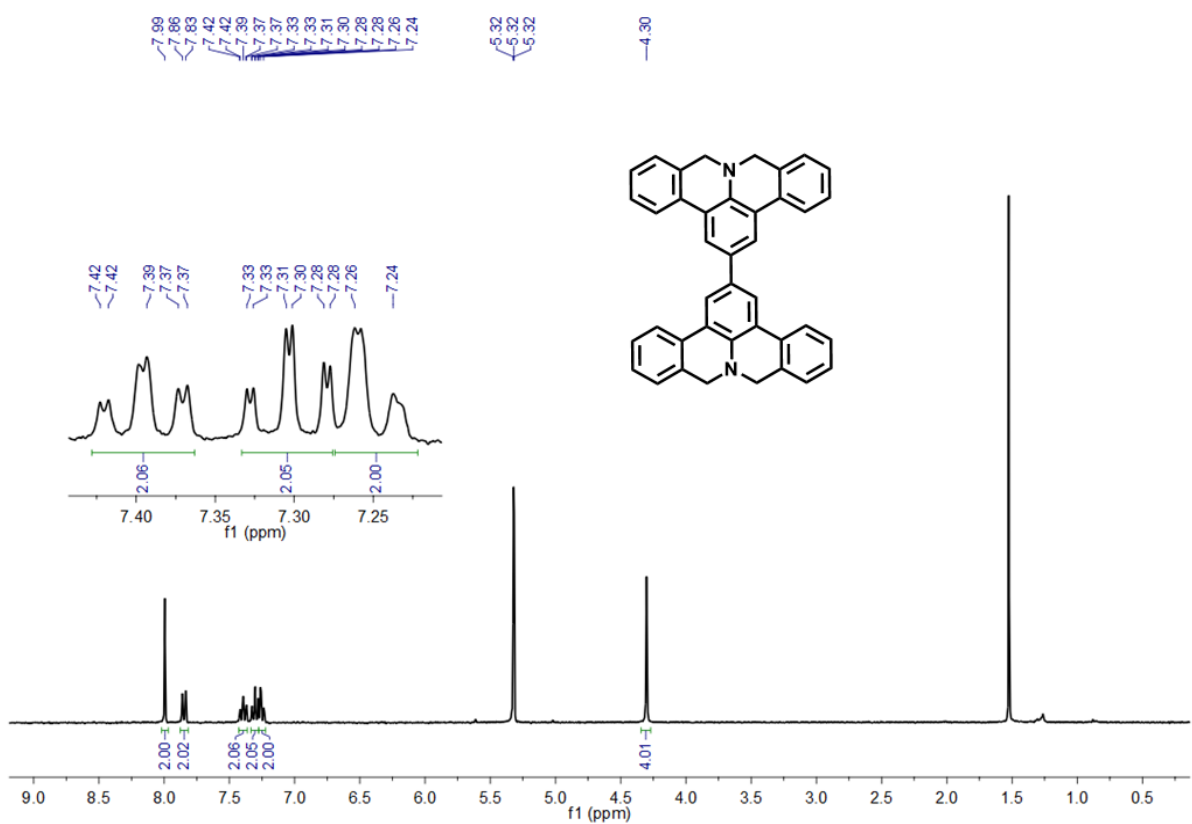

**Figure S5.** $^1$H NMR spectrum (300 MHz) of **3** in CD$_2$Cl$_2$ at room temperature.

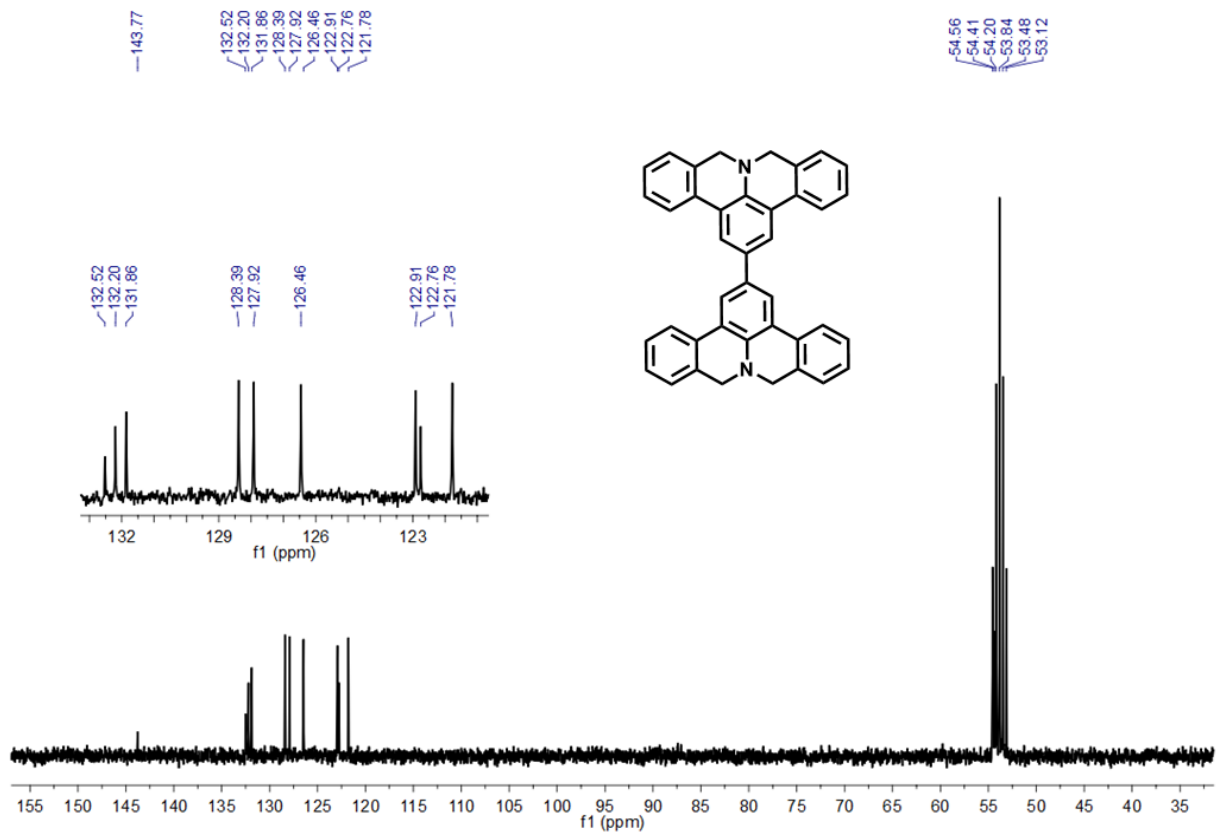

**Figure S6.** $^{13}$C NMR spectrum (75 MHz) of **3** in CD$_2$Cl$_2$ at room temperature.

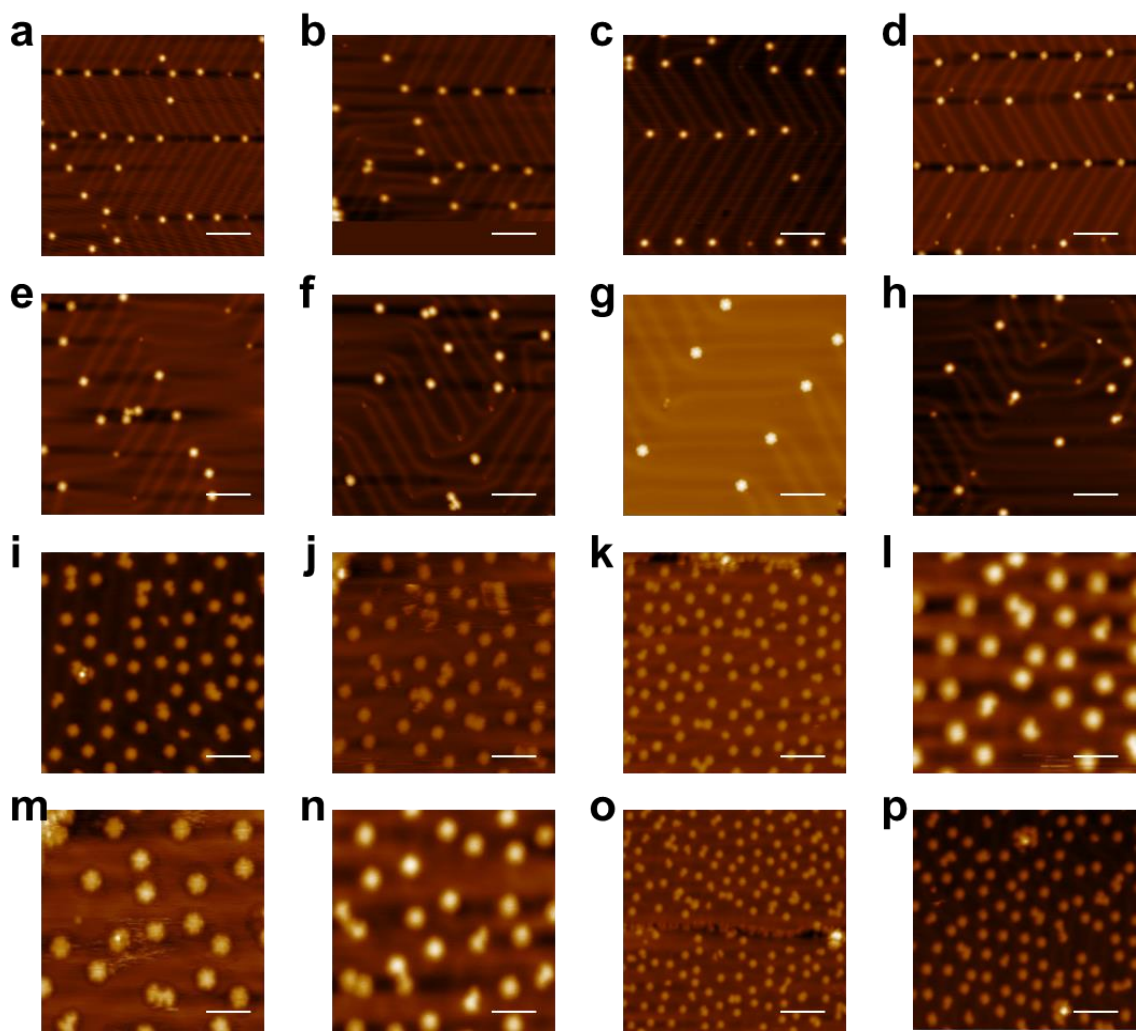

**Figure S7. STM images of on-surface synthesized diaza-HBC monomers by homocoupling of 1.** (**a**) - (**p**), STM images used for the statistics. (Scanning parameters: **a:** V = -0.4 V, I= 5 pA, **b:** V = -0.5 V, I= 100 pA, **c:** V = -0.5 V, I= 100 pA, **d:** V = -1 V, I= 100 pA, **e:** V = -0.6 V, I= 100 pA, **f:** V = -1 V, I= 1 nA, **g:** V = -0.5 V, I= 100 pA, **h:** V = -0.6 V, I= 100 pA, **i:** V = -0.1 V, I= 10 pA, **j:** V = -0.1 V, I= 10 pA, **k:** V = -0.04 V, I= 10 pA, **l:** V = -0.04 V, I= 10 pA, **m:** V = -0.1 V, I= 10 pA, **n:** V = -0.1 V, I= 10 pA, **o:** V = -0.1 V, I= 10 pA, **p:** V = -0.04 V, I= 10 pA, ). Scale bars: (**a**) 12 nm, (**b**, **c**, **d**, **o**) 10 nm, (**e**, **f**, **h**, **k**, **p**) 8 nm, (**g**, **i**, **j**) 6 nm, (**l**, **m**, **n**) 4 nm.

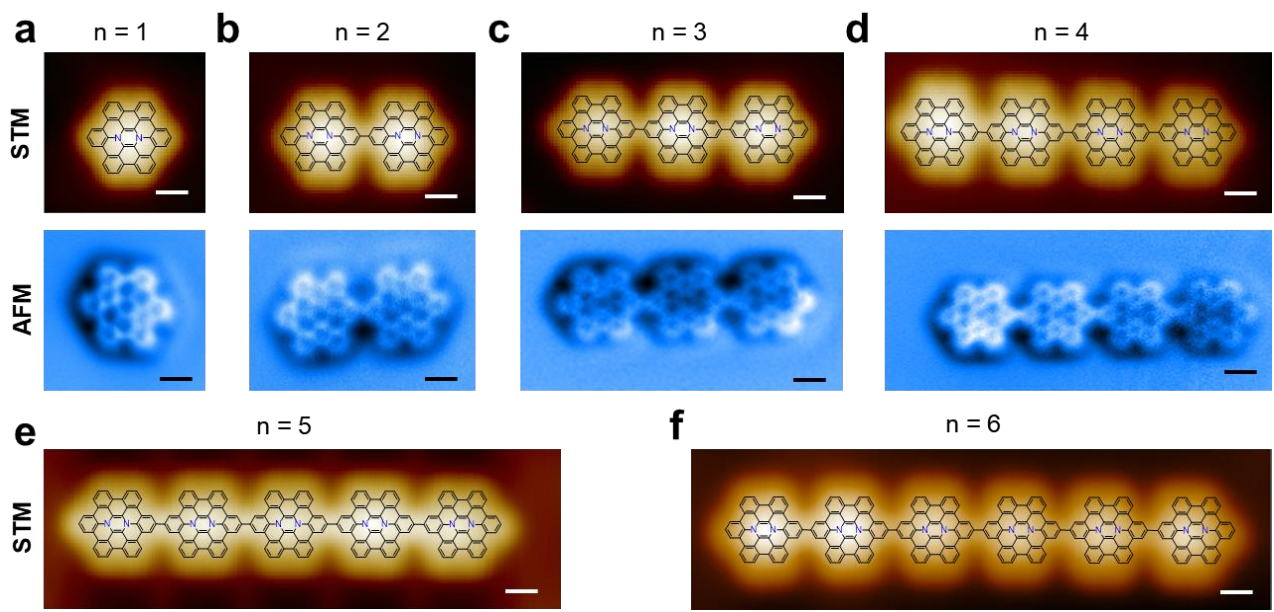

**Figure S8. STM and nc-AFM images of chain 1 with n = 1 - 6.** (**a**) - (**d**) Upper panels: STM images of **chain 1** with n = 1 - 4 (V = -0.1 V, I = 100 pA); lower panels: Chemical-bond-resolved constant-height nc-AFM images of the respective **chain 1** in the upper panels. The images were obtained with a CO-functionalized tip. (oscillation amplitude $A_{osc}$ = 100 pm). (**e**) and (**f**) STM images of **chain 1** with n = 5 and n = 6, respectively (**e:** V = -0.1 V, I = 100 pA; **f:** V = -0.9 V, I = 1 nA). Scale bars: 0.5 nm.

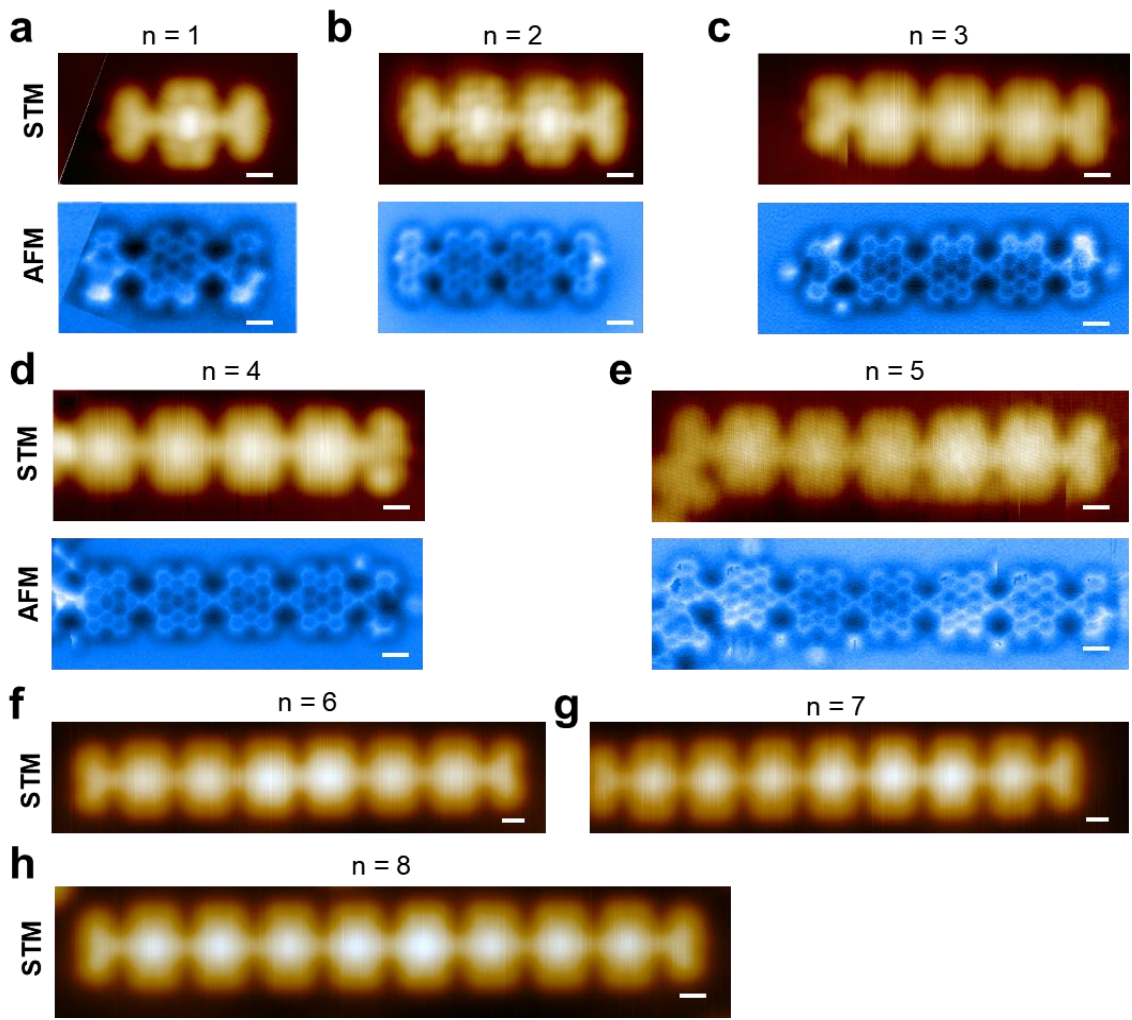

**Figure S9. STM and nc-AFM images of chain 2 with n = 1 - 8.** (**a**) - (**e**) Upper panels: STM images of **chain 2** with n = 1 - 5 (V = -0.1 V, I = 100 pA); lower panels: Chemical-bond-resolved constant-height nc-AFM images of the respective **chain 2** in the upper panels. The images were obtained with a CO-functionalized tip. (oscillation amplitude $A_{OSC}$ = 100 pm). (**f**) - (**h**), STM images of **chain 2** with n = 6 - 8 (**f:** V = -0.9 V, I = 1 nA; **g:** V = -0.1 V, I = 1 nA; **h:** V = -0.1 V, I = 1 nA). Scale bars: 0.5 nm.

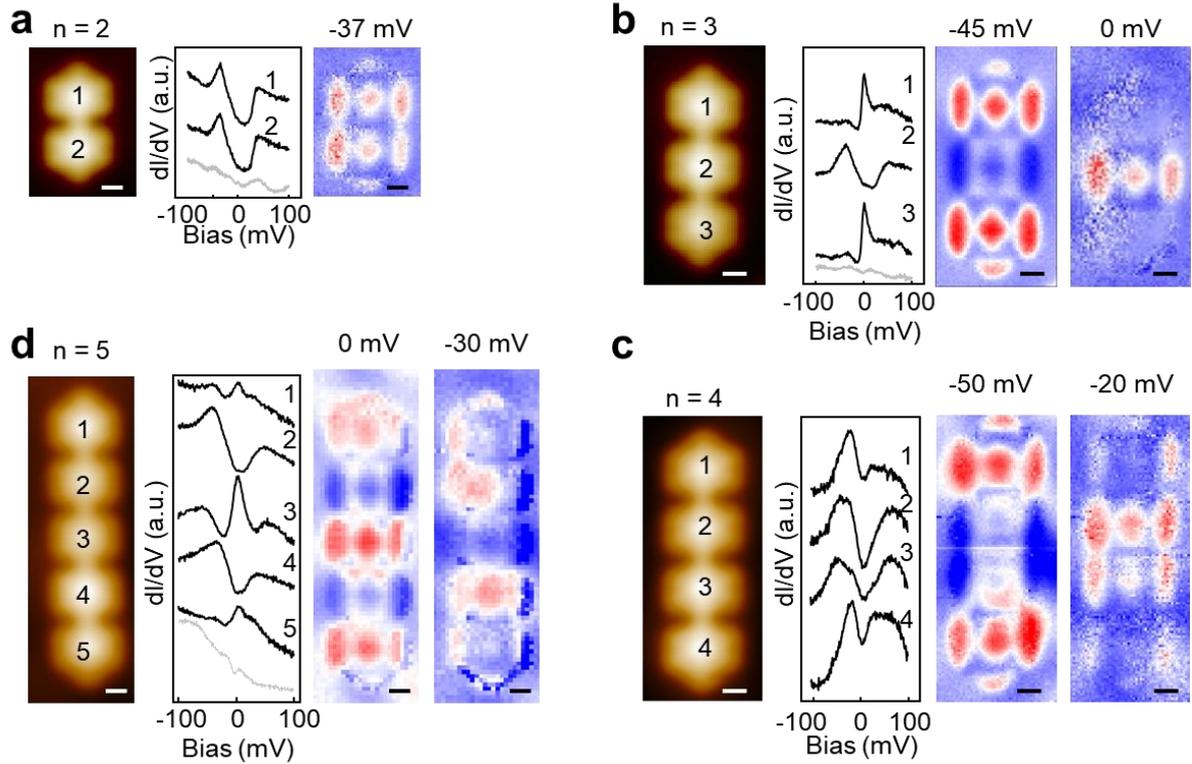

**Figure S10. Spin excitations in chain 1 with n = 2 - 5.** (**a**) - (**d**) The STM images, dI/dV spectra, and dI/dV maps of **chain 1** with n = 2 – 5. dI/dV spectra were taken at the center of each unit ($V_{AC}$ = 1 mV). dI/dV maps of diaza-HBC chains were taken at energies marked at the top of each map. STM scanning parameters: (**a**) – (**d**) V = -0.1 V, I = 1 nA. dI/dV map parameters: V = -0.1 V, $V_{AC}$ = 10 mV for the maps in (a) - (d). Scale bars: 0.5 nm.

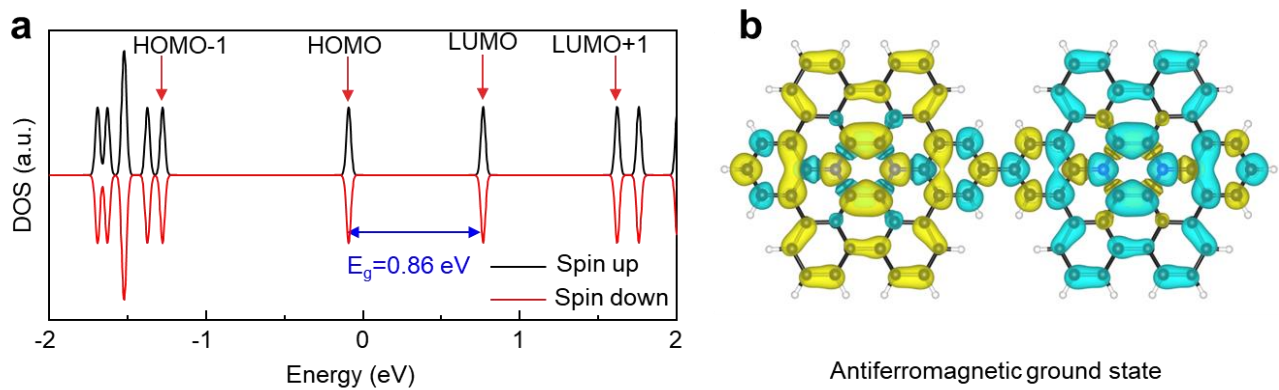

**Figure S11. DFT calculated ground state of the diaza-HBC dimer subtracting one electron at each unit.** (**a**) Total spin DOS of diaza-HBC dimer. The black and red curves represent spin up and spin down DOS, respectively. (**b**) Spin-polarized electron density distribution of the diaza-HBC dimer, showing an antiferromagnetic ground state.

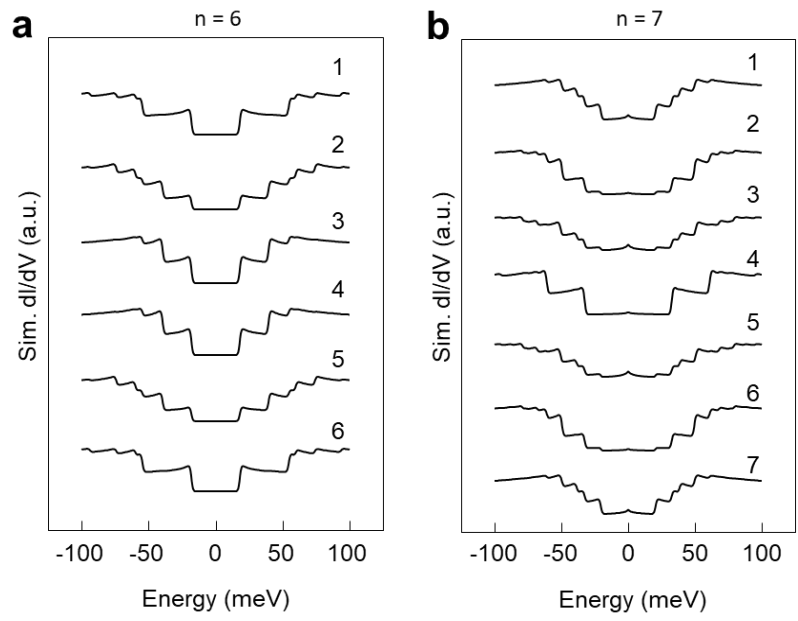

**Figure S12. Theoretical calculated dI/dV spectra on n = 6 and 7 chains based on *S* = 1/2 antiferromagnetic Heisenberg model chain with an effective temperature of 5 K.**

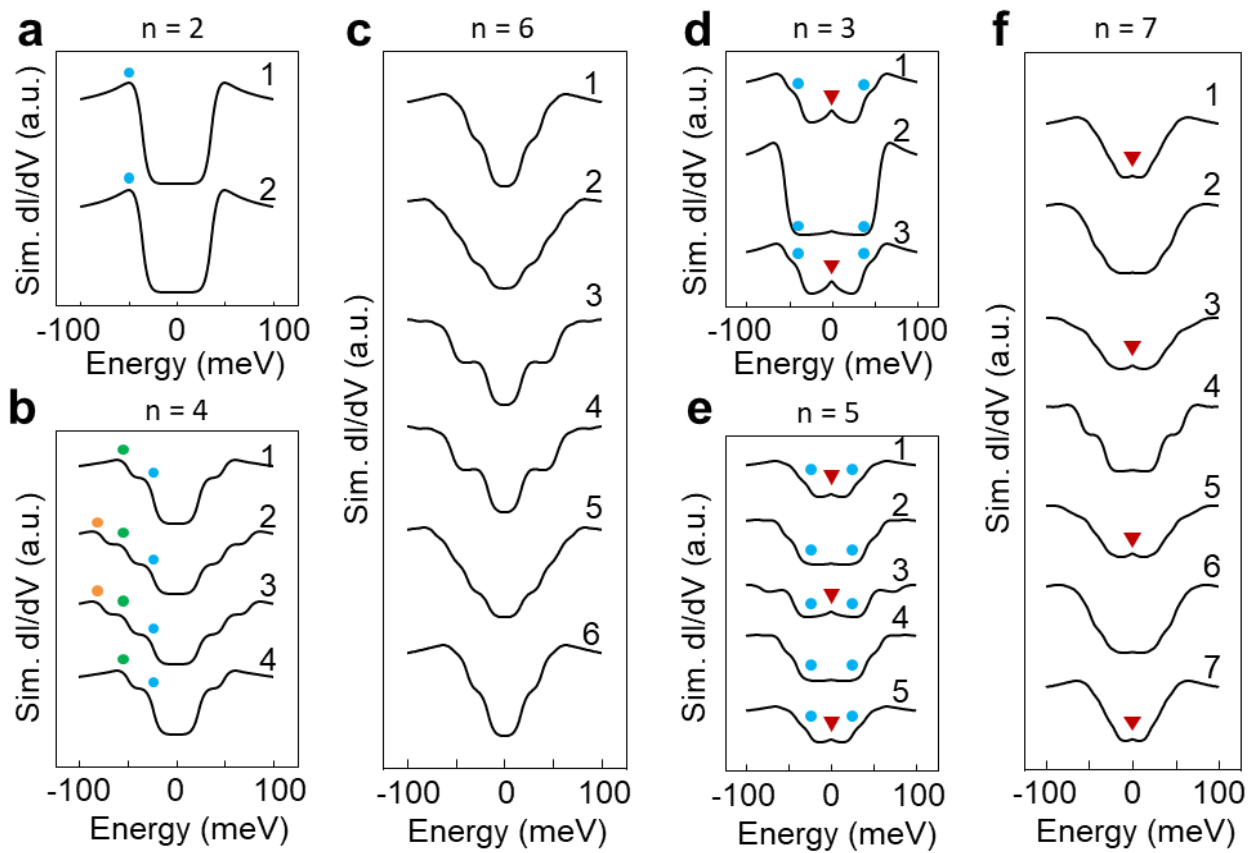

**Figure S13. Theoretical calculated dI/dV spectra on n =2-7 chains based on *S* = 1/2 antiferromagnetic Heisenberg model chain with an effective temperature of 30 K.** The blue, green, and orange dots denote the first, second and third spin excitations, respectively. The red triangles denote the enhanced Kondo resonance.